\documentclass[reprint]{JASA}

\usepackage[caption=false,font=footnotesize]{subfig}
\usepackage{algpseudocode}
\usepackage{booktabs}
\usepackage{bm}
\usepackage{multirow}
\usepackage{tikz}

\begin{document}

\title[Deep Sound Field Reconstruction in Real Rooms: Introducing the ISOBEL Sound Field Dataset]{Deep Sound Field Reconstruction in Real Rooms: Introducing the ISOBEL Sound Field Dataset}
\author{Miklas Str{\o}m Kristoffersen}
\affiliation{Research Department, Bang \& Olufsen a/s, Struer, Denmark}
\affiliation{AI and Sound Section, Department of Electronic Systems, Aalborg University, Aalborg, Denmark}
\author{Martin Bo M{\o}ller}
\author{Pablo Mart\'inez-Nuevo}
\affiliation{Research Department, Bang \& Olufsen a/s, Struer, Denmark}

\author{Jan {\O}stergaard}
\affiliation{AI and Sound Section, Department of Electronic Systems, Aalborg University, Aalborg, Denmark}


\date{\today}

\begin{abstract}
Knowledge of loudspeaker responses are useful in a number of applications, where a sound system is located inside a room that alters the listening experience depending on position within the room.
Acquisition of sound fields for sound sources located in reverberant rooms can be achieved through labor intensive measurements of impulse response functions covering the room, or alternatively by means of reconstruction methods which can potentially require significantly fewer measurements.
This paper extends evaluations of sound field reconstruction at low frequencies by introducing a dataset with measurements from four real rooms.
The ISOBEL Sound Field dataset is publicly available, and aims to bridge the gap between synthetic and real-world sound fields in rectangular rooms.
Moreover, the paper advances on a recent deep learning-based method for sound field reconstruction using a very low number of microphones, and proposes an approach for modeling both magnitude and phase response in a U-Net-like neural network architecture.
The complex-valued sound field reconstruction demonstrates that the estimated room transfer functions are of high enough accuracy to allow for personalized sound zones with contrast ratios comparable to ideal room transfer functions using 15 microphones below 150 Hz.
\end{abstract}


\maketitle


\section{\label{sec:1} Introduction}

The response of a sound system in a room primarily varies with the room itself, the position of the loudspeakers, and the listening position. 
In order to deliver the intended sound system behavior to listeners, it is necessary to know about and compensate for this effect. 
Applications include among others room equalization \citep{Cecchi2018,Karjalainen2001,Radlovic2000}, virtual reality sound field navigation \citep{Tylka2015}, source localization \citep{Nowakowski2017}, and spatial sound field reproduction over predefined or dynamic regions of space also referred to as sound zones \citep{Betlehem2015,Moller2020}.
An approach to achieve this, is to measure the loudspeaker response at the desired listening locations and adjust the sound system accordingly.
However, the task of measuring impulse responses on a sufficiently fine-grained grid in an entire room, quickly poses as a time-consuming and extensive manual labor that is not desirable.
Instead, methods have been developed for the purpose of estimating impulse responses in a room based on a limited number of actual measurements.
These methods are also referred to as sound field reconstruction and virtual microphones.
The task of reconstructing room impulse responses in positions that have not been measured directly, is an active research field which has been explored in several studies \citep{Ajdler2006,Mignot2014,Antonello2017,Verburg2018,Fernandez-grande2019,Vu2020}.

Machine learning, and in particular deep learning, is currently receiving widespread attention across scientific domains, and as an example within room acoustics, it has been used to estimate acoustical parameters of rooms \citep{Genovese2019,Yu2021}.
In recent work, deep learning-based methods were introduced to sound field reconstruction in reverberant rectangular rooms \citep{Lluis2020}.
This data-driven approach is able to learn sound field magnitude characteristics from large scale volumes of simulated data without prior information of room characteristics, such as room dimensions and reverberation time.
The method is computationally efficient, and works with irregularly and arbitrarily distributed microphones for which there is no requirement of knowing absolute locations in the Euclidean space, in contrast to previous solutions.
Furthermore, the reconstruction proves to work with a very low number of microphones, making real-world implementation feasible.
To assess the issue of real-world sound field reconstruction, the method is evaluated using measurements in a single room \citep{Lluis2020}.
However, it is still unknown how much knowledge is transferred from the simulated to the real environment, as well as how well the model generalizes to different real rooms.
This is a general problem in deep learning applications that rely on labor intensive data collections, which is our motivation for publishing an open access dataset of real-world sound fields in a diverse set of rooms.

This paper studies sound field reconstruction at low frequencies in rectangular rooms with a low number of microphones.
The main contributions are:
\begin{itemize}
    \item This paper introduces a sound field dataset, which is publicly available for development and evaluation of sound field reconstruction methods in four real rooms. It is our hope that the ISOBEL Sound Field dataset will help the community in benchmarking and comparing state-of-the-art results.
    \item We assess the real-world performance of deep learning-based sound field magnitude reconstruction trained on simulated sound fields. For this purpose, we consider low frequencies, since low-frequency room modes can significantly alter listening experience.
        Furthermore, we are interested in using a very low number of microphones.
    \item Moreover, we extend the deep learning-based sound field reconstruction to cover complex-valued inputs, i.e. both the magnitude and the phase of a sound field. Evaluation is performed in both simulated and real rooms, where a performance gap is observed. We argue why complex sound field reconstruction may have more difficulties in transferring useful knowledge from synthetic to real data.
    \item Lastly, we demonstrate the application of complex-valued sound field reconstruction within the field of sound zone control. Specifically, it is shown that sound fields reconstructed from as little as five microphones pose as valuable inputs to acoustic contrast control.
\end{itemize}

The paper is organized as follows: Section~\ref{sec:sf_recon} introduces the concept of sound field reconstruction.
Details of measurements from real rooms are presented in Section~\ref{sec:dataset}.
In Section~\ref{sec:mag}, we focus on the problem of reconstructing the magnitude of sound fields, while Section~\ref{sec:complex} extends the model to complex-valued sound fields.
Finally, Section~\ref{sec:sound_zones} investigates the application of sound zones through sound field reconstruction.
\section{Sound Field Reconstruction}\label{sec:sf_recon}
Our approach towards the sound field reconstruction problem is based on the observation that acoustic pressure in a room can be described using a three-dimensional regular grid of points defining a three-dimensional discrete function.
The approach specifically for the purpose of magnitude reconstruction was introduced in~\citep{Lluis2020}.
First, let $\mathcal{R}=[0,l_x]\times[0,l_y]\times[0,l_z]$ denote a rectangular room, where $l_x,l_y,l_z>0$ are the length, width, and height of the room, respectively.
Given such room, we define the grid as a discrete set of coordinates $\mathcal{D}_o$. 
However, for the sake of simplicity, we reduce the three-dimensional problem to a two-dimensional reconstruction on horizontal planes.
The two-dimensional grid with a constant height $z_o$ is defined as
\begin{equation}
\mathcal{D}_o:=\Big\{\Big(i\frac{l_x}{I-1},j\frac{l_y}{J-1},z_o\Big)\Big\}_{i,j}
\end{equation}
for $z_o\in[0,l_z]$, $i=0,\ldots,I-1$, $j=0,\ldots,J-1$, and integers $I,J\geq2$.
Note, though, that the dataset collected for this study, which we will introduce in Section~\ref{sec:dataset}, does in fact contain multiple horizontal planes at different heights.
We keep the investigations of three-dimensional reconstruction for future work, and frame the core challenge of this paper as estimation of sound pressure in two-dimensional horizontal planes.

The function that we seek to reconstruct on this grid is the Fourier transform of the sound field in a frequency band that covers the low frequencies.
The complex-valued frequency-domain sound field calculated using the Fourier transform is given by
\begin{equation}
s(\mathbf{r},\omega):=\int_\mathbb{R}p(\mathbf{r},t)e^{-j\omega t}\mathrm{d}t\label{eq:2}
\end{equation}
where $\omega\in\mathbb{R}$ is a given excitation frequency, and $p(\mathbf{r},t)$ denotes the spatio-temporal sound field with $\mathbf{r}\in\mathcal{R}$.
We refer to the real and imaginary parts of the sound field using $s_\text{Re}(\mathbf{r},\omega)$ and $s_\text{Im}(\mathbf{r},\omega)$, respectively.
Note that $s$ is defined as the magnitude of the Fourier transform in~\citep{Lluis2020}.
Instead, for magnitude reconstruction, we introduce the magnitude of the sound field
\begin{equation}
\left|s(\mathbf{r},\omega)\right|:=\left|\int_\mathbb{R}p(\mathbf{r},t)e^{-j\omega t}\mathrm{d}t\right|
\end{equation}
for $\omega\in\mathbb{R}$ and $\mathbf{r}\in\mathcal{R}$.

The procedure for reconstructing $s(\mathbf{r},\omega)$ on $\mathcal{D}_o$ takes its starting point from actual observations of the sound field in select positions of the grid.
We refer to the collected set of these available sample points as $\mathcal{S}_o$, which we further define to be a subset of the full grid.
That is, $\mathcal{S}_o\subseteq\mathcal{D}_o$.
The cardinality $|\mathcal{S}_o|$ of the set $\mathcal{S}_o$ is the number of available sample points, which we will also refer to as the number of microphones $n_{mic}$ in later experiments.
We define the samples available to the reconstruction algorithm as
\begin{equation}
\{s(\mathbf{r},\omega)\}_{\mathbf{r}\in\mathcal{S}_o\subseteq\mathcal{D}_o}.
\end{equation}

An important aspect of these definitions is that the grid is unitless and positions can be defined in relative terms.
That is, when sampling a point in the grid, only the relative position within the grid, and hence the room, needs to be known.
This allows us to relax the data collection compared to alternative methods that require absolute locations.
Another important element to consider is that the sampling pattern of $\mathcal{S}_o$ can form any arrangement within $\mathcal{D}_o$ as long as $1\leq|\mathcal{S}_o|\leq|\mathcal{D}_o|$.
As an example, this means that sampled points can be irregularly distributed spatially in a room.

Situations may arise where the sound field resolution, as defined by $l_x$, $I$, $l_y$, and $J$, is too coarse.
As an example, consider rooms that are either very long, wide, or in general large.
Another example includes applications where fine-grained variations within a sound field are of importance.
To compensate for this effect, we allow the reconstruction to base its output on another grid than $\mathcal{D}_o$.
Such domain will typically be an upsampling of the original grid, but similarly it can be defined with other transformations, e.g. downsampling.
Specifically, we define the grid as
\begin{equation}
\mathcal{D}_o^{L,P}:=\Big\{\Big(i\frac{l_x}{IL-1},j\frac{l_y}{JP-1},z_o\Big)\Big\}_{i,j}
\end{equation}
where $i=0,\ldots,IL-1$, $j=0,\ldots,JP-1$, and $L,P$ must be chosen such that $IL,JP\in\mathbb{Z}^+$.
Note that a value larger than one for either $L$ or $P$ results in an upsampling in the respective dimension.

The task of the sound field reconstruction is then to estimate the sound field on the grid $\mathcal{D}_o^{L,P}$ based on the sampled points $\mathcal{S}_o$.
In particular, the objective of the reconstruction algorithm is to learn parameters $\mathbf{w}$ given
\begin{eqnarray}
g_\mathbf{w}:&\ \mathbb{C}^{|\mathcal{S}_o|K}&\ \to\ \mathbb{C}^{|\mathcal{D}_o^{L,P}|K}\\\nonumber
&\{s(\mathbf{r},\omega_k)\}_{\mathbf{r}\in\mathcal{S}_o,\omega_k\in \Omega}&\ \mapsto\ \{\hat{s}(\mathbf{r},\omega_k)\}_{\mathbf{r}\in\mathcal{D}_o^{L,P},\omega_k\in \Omega}
\end{eqnarray}
where $g_\mathbf{w}$ is an estimator and $\Omega=\{\omega_k\}_{k=1}^K$ is the set of frequencies at which the sound field will be reconstructed.
The remainder of the paper describes the procedure for learning parameters $\mathbf{w}$ using deep learning-based methods.

\begin{figure*}[tb]
    \centering%
    \begin{minipage}{0.17\textwidth}%
    \includegraphics[height=8.12cm]{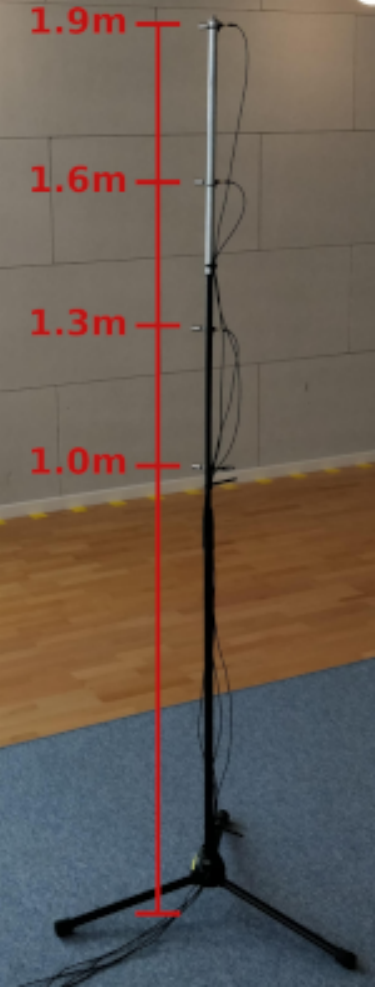}\hfill
    \end{minipage}%
    \begin{minipage}{0.32\textwidth}%
        \includegraphics[height=4cm]{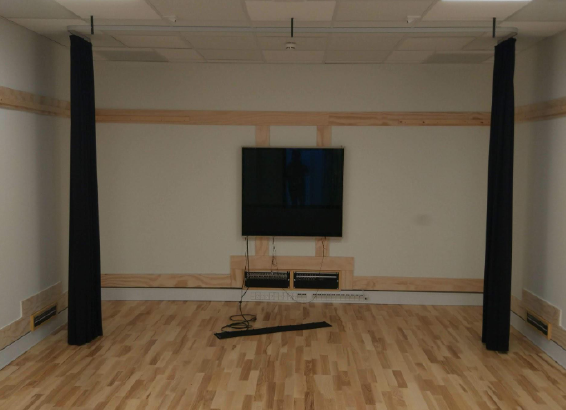}\\\vspace{0.06cm}
    \includegraphics[height=4cm]{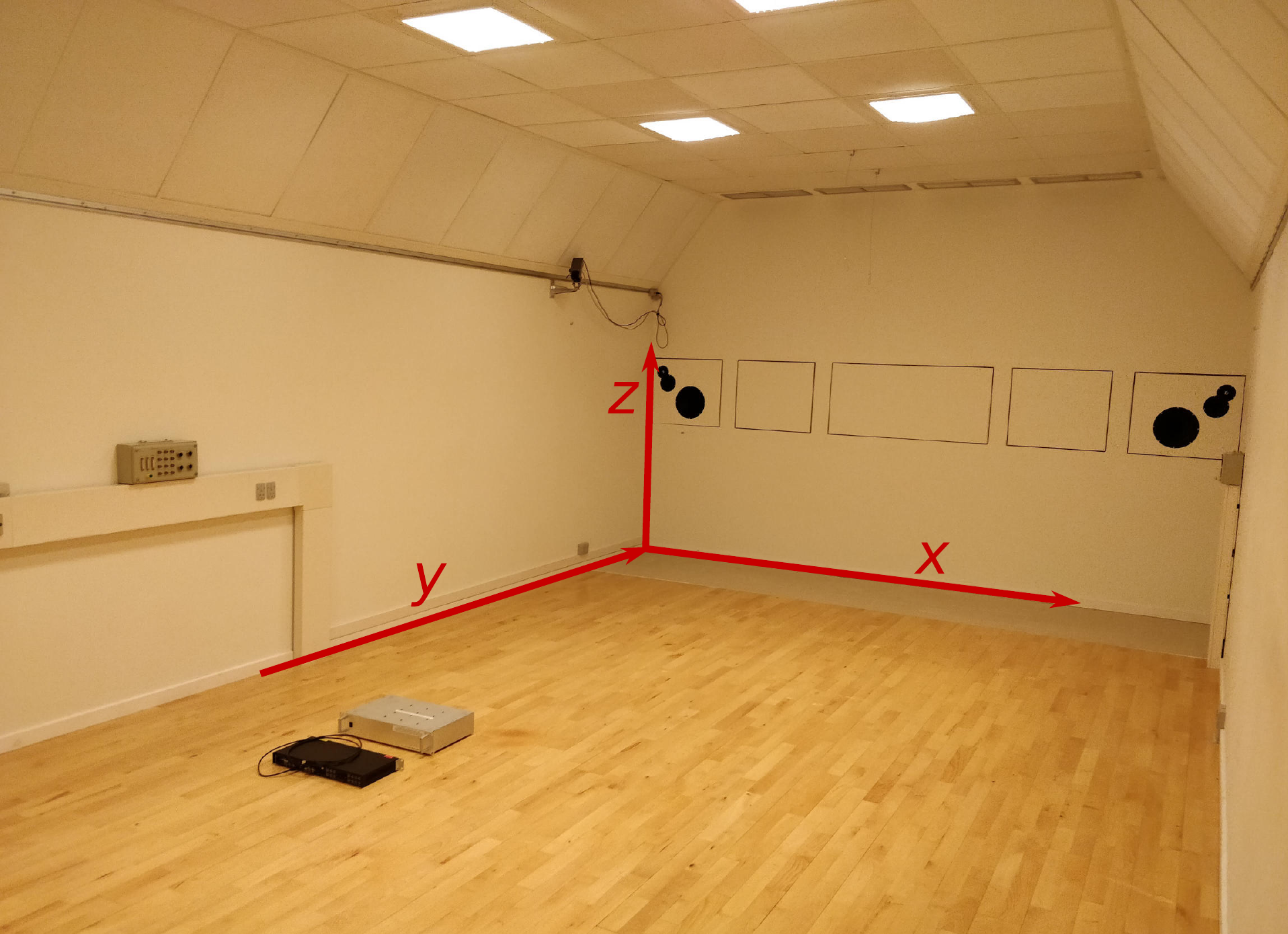}
    \end{minipage}%
    \begin{minipage}{0.30\textwidth}%
        \includegraphics[height=4cm]{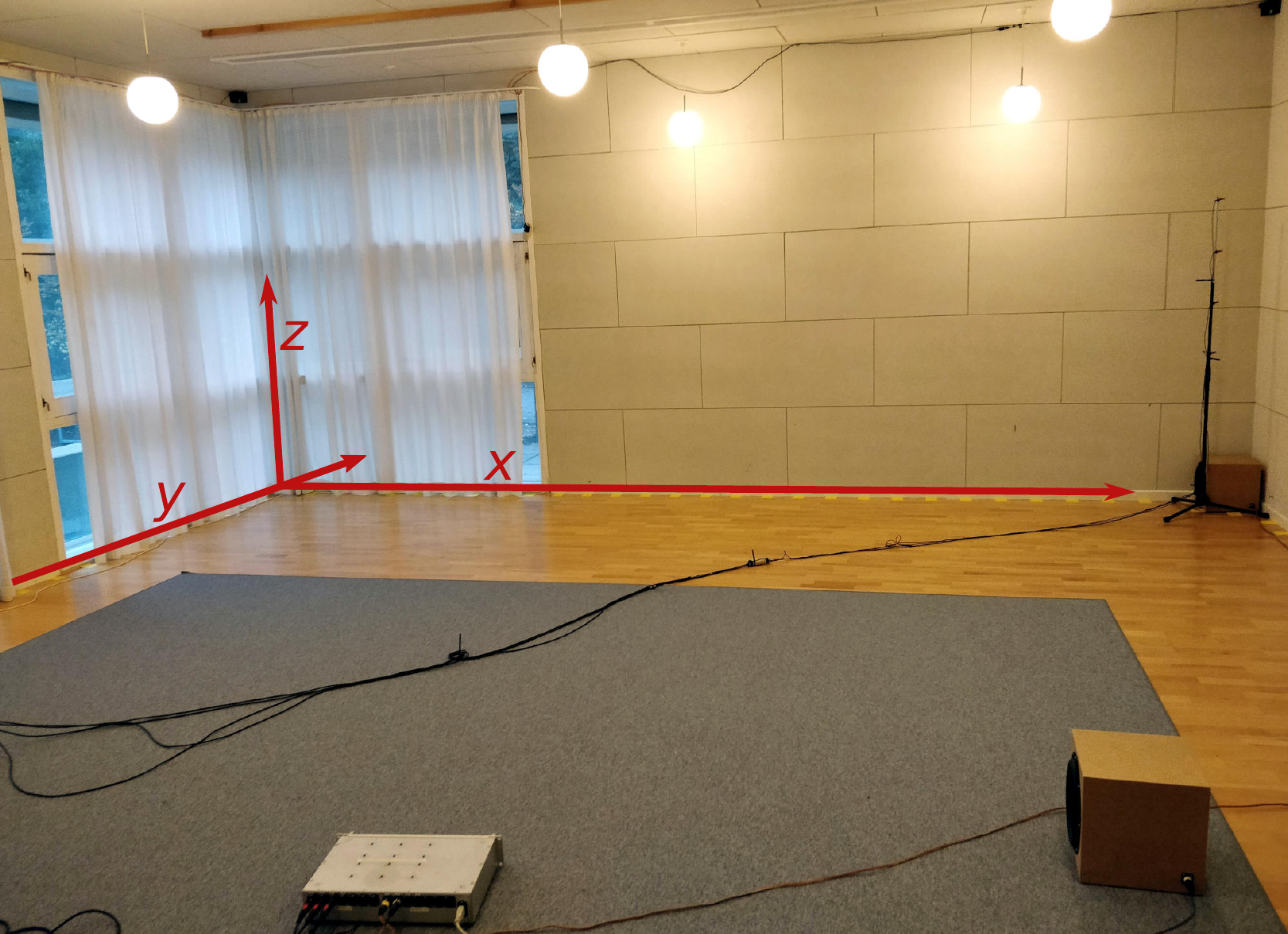}\\\vspace{0.06cm}
    \includegraphics[height=4cm]{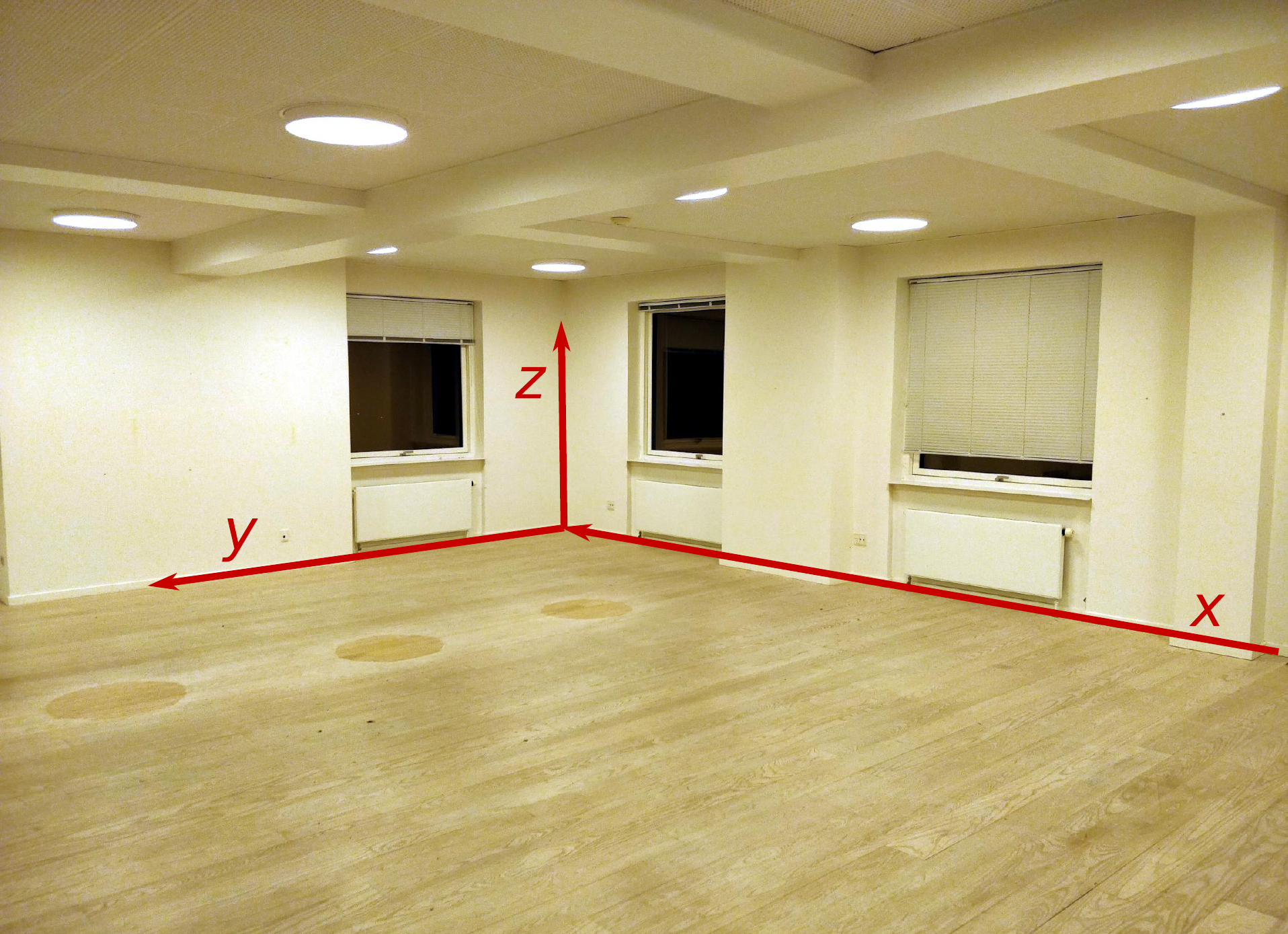}
    \end{minipage}%
\caption{Left: Rig with four microphones. Rooms from top left to bottom right: Room B, VR Lab, Listening Room, and Product Room.}%
\label{fig:mic_rooms}%
\end{figure*}

\subsection{Evaluation Metrics}
The successfulness of the estimator is quantitatively judged using normalized mean square error (NMSE) at each frequency point in $\{\omega_k\}_{k=1}^K$
\begin{equation}
    \text{NMSE}_k=\frac{\sum\limits_{\mathbf{r}\in\mathcal{D}_o^{L,P}}|s(\mathbf{r},\omega_k)-\hat{s}(\mathbf{r},\omega_k)|^2}{\sum\limits_{\mathbf{r}\in\mathcal{D}_o^{L,P}}|s(\mathbf{r},\omega_k)|^2}.
\end{equation}
The NMSE provides an average error over all positions in the grid between reconstructed and original sound fields for a single room at a single frequency.
We also introduce an average NMSE, which is the NMSE performance averaged over all frequencies of interest as well as over all realizations from $M$ trials, e.g. multiple rooms
\begin{align}
    &\text{MNMSE}=\nonumber\\
    &\frac{1}{MK}\sum^{M}_{m=1}\sum^{K}_{k=1}\frac{\sum\limits_{\mathbf{r}\in\mathcal{D}_o^{L,P}}|s_m(\mathbf{r},\omega_k)-\hat{s}_m(\mathbf{r},\omega_k)|^2}{\sum\limits_{\mathbf{r}\in\mathcal{D}_o^{L,P}}|s_m(\mathbf{r},\omega_k)|^2}.
\end{align}
This measure serves as an overall indication of the accuracy of a model, whereas the NMSE$_k$ allows a deeper insight of model behaviors at different frequencies.
Note that the $M$ trials are specific to each experiment and will be described accordingly.
\section{The ISOBEL Sound Field Dataset}\label{sec:dataset}
A major contribution of this paper is the ISOBEL Sound Field dataset, which is released as open access alongside the manuscript.\footnote{The data are collected under the Interactive Sound Zones for Better Living (ISOBEL) project, which aims to develop interactive sound zone systems, responding to the need for sound exposure control in dynamic real-world contexts, adapted to and tested in healthcare and homes. The ISOBEL Sound Field dataset can be accessed at \url{https://doi.org/10.5281/zenodo.4501339}.}
The intended purpose is to use the measurements from real rooms for evaluation of sound field reconstruction in a diverse set of rooms.
Note that the room-wide measurements of room impulse responses have several other use-cases that will not be further investigated in this paper, but we encourage the use outside sound field reconstruction as well.
This section details the dataset and the measurement procedure.

\begin{table}[b]
    \caption{Room characteristics in the ISOBEL Sound Field dataset.
The reverberation times are the arithmetic averages of 1/3 octave $T_{20}$ estimates in the frequency range 50-316Hz.
    }
    \label{tab:dataset}
\centering
\newcolumntype{P}[1]{>{\centering\arraybackslash}p{#1}}
    \begin{tabular}{lP{2.8cm}P{2.1cm}P{1.3cm}}
    \toprule
        \textbf{Room}    & \textbf{Dim. [m]} & \textbf{Size [m$^2$/m$^3$]}  & \textbf{$\bm{T_{20}}$ [s]}\\
    \midrule
    Room B           & 4.16 x \phantom{1}6.46 x 2.30 & \phantom{1}27/\phantom{1}62  & 0.39\\
    VR Lab           & 6.98 x \phantom{1}8.12 x 3.03 & \phantom{1}57/172 & 0.37\\
    List. Room       & 4.14 x \phantom{1}7.80 x 2.78 & \phantom{1}32/\phantom{1}90  & 0.80\\
    Prod. Room       & 9.13 x 12.03 x 2.60           & 110/286  & 0.77\\
    \bottomrule
\end{tabular}
\end{table}
The dataset consists of measurements from four different rooms as specified in Table~\ref{tab:dataset} and depicted in Fig.~\ref{fig:mic_rooms}.
The data collection is an extension to the real room measured in~\citep{Lluis2020}, which is included in the ISOBEL Sound Field dataset as Room B for simple access to all measured rooms.
The rooms are located at Aalborg University, Aalborg, Denmark, and Bang \& Olufsen a/s, Struer, Denmark.
The rooms have significantly different acoustic properties and also vary in size.
Two types of measurements are conducted in each room: 1)~Reverberation time; 2)~Sound field.
However, only the sound field measurements are released as part of the dataset.

The reverberation times are measured in conformity with ISO 3382-2 \citep{ISO3382} and calculated based on resulting impulse responses using backwards integration and least-squares best fit evaluation of the decay curves.\footnote{\label{note1}Further details of the experimental setup and protocol, e.g. equipment, are available in the measurement reports included with the dataset.}
The reverberation times reported in the table are the arithmetic averages of 1/3 octave $T_{20}$ estimates in the frequency range 50-316 Hz.

The sound field measurements are performed on a 32 by 32 grid with sample points distributed uniformly along the length and width of each room.
That is, a total of 1024 positions are measured in each room if possible, but in some cases it is not feasible to measure all positions due to e.g. obstacles.\footnote{See footnote \ref{note1}.}
The horizontal grids are measured at four different heights: 1, 1.3, 1.6, and 1.9 meters above the floor.\footnote{Room B has measurements at a single height: 1 meter above the floor.}
This is achieved using the microphone rig depicted in Fig.~\ref{fig:mic_rooms}.
Two 10 inch loudspeakers are used to acquire sound fields from two different source positions in each room.
Both loudspeakers are placed on the floor, one in a corner and one in an arbitrary position.
The sound sources are kept in the same position, while the microphones are moved around the room to record impulse responses.
For each microphone position in the grid, the two sources play logarithmic sine sweeps in the frequency range 0.1-24,000~Hz followed by a quiet tail, \citep{Farina2000}.
We use a sampling frequency of 48,000~Hz. 
The equipment includes among others four G.R.A.S. 40AZ prepolarized free-field microphones connected to four G.R.A.S. 26CC CCP standard preamplifiers and an RME Fireface UFX+ sound card. 
The four microphones are level calibrated at 1,000 Hz using a Br{\"u}el \& Kj{\ae}r sound calibrator type 4231 prior to the measurements.
\section{Sound Field Magnitude Reconstruction}\label{sec:mag}
In the previous sections we have introduced the problem of reconstructing sound fields on two-dimensional grids in rectangular rooms, as well as introduced a real-world dataset specifically for evaluation of estimators solving such problem.
In recent work, \citep{Lluis2020} showed that the problem fits within the context of deep learning-based methods for image reconstruction.
Specifically, the tasks of inpainting, \citep{Bertalmio2000,Liu2018}, and super-resolution, \citep{Dong2016,Ledig2017}, which can be paralleled to the tasks of filling in the grid points that are not measured in the sound fields $\mathcal{D}_o^{L,P} \setminus \mathcal{S}_o$, as well as upsampling the grid resolution to achieve fine-grained variations in sound fields.
One realization is that these methods are designed to work with real-valued images.
To accommodate this, \citep{Lluis2020} propose to reconstruct only the magnitude of the sound field, i.e. $|s(\mathbf{r},\omega)|$, using a U-Net-like architecture, \citep{Ronneberger2015}. 

To this end, the sampled grids are defined as tensors together with masks specifying which positions are measured \citep{Lluis2020}.
As an example, $\{|s(\mathbf{r},\omega_k)|\}_{\mathbf{r}\in\mathcal{D}_o^{L,P},k}$ can be constructed as a tensor of the form $\mathbf{S}_{mag}\in\mathbb{R}^{IL \times JP \times K}$.
The network is trained using a large number of simulated realizations of rooms, as will be described in the following section.
For the experiments, we are interested in assessing the ability of the model to generalize to a wide range of real rooms.

\subsection{Simulation of Sound Fields for Training Data}\label{sec:sim_data}
Green's function can be used to approximate sound fields in rectangular rooms that are lightly damped, \citep{Jacobsen2013}.
The function provides a solution as an infinite summation of room modes in the three dimensions of a room, $x$, $y$, and $z$.
It is defined as follows
\begin{equation}
    G(\mathbf r,\mathbf r_0, \omega) \approx -\frac{1}{V}\sum_{N} \frac{\psi_{N}(\mathbf r) \psi_{N}(\mathbf r_0)}{(\omega/c)^2 - (\omega_{N}/c)^2 -j\omega /\tau_N}
\end{equation}
where $\sum_N = \sum_{n_x=0}^\infty \sum_{n_y=0}^\infty \sum_{n_z=0}^\infty$, for compactness, denotes summation across modal orders in the three dimensions of the room, and similarly the triplet of integers $(n_x,n_y,n_z)$ are represented by $N$.
Furthermore, $V$ denotes the volume of the room, $\omega_{N}^2$ represents angular resonance frequency of a mode associated with a specific $N$, the shape of the mode is denoted $\psi_N(\cdot)$, $\tau_N$ is the time constant of the mode, and $c$ is the speed of sound.
Assuming rigid boundaries, the shape is determined using the expression \citep{Jacobsen2013}
\begin{equation}
    \psi_N(\mathbf x) = \Lambda_N \cos \frac{n_x \pi x}{l_x} \cos \frac{n_y \pi y}{l_y} \cos \frac{n_z \pi z}{l_z}.
\end{equation}
Here, $\Lambda_N = \sqrt{\epsilon_x \epsilon_y \epsilon_z}$ are constants used for normalization with $\epsilon_0 = 1$, $\epsilon_1 = \epsilon_2 = \cdots = 2$.
Using Sabine's equation, the absorption coefficient is calculated and used to determine time constants of each mode.
This is done by assuming that surfaces of a room have uniform distribution of absorption.

In the following experiments, two sets of training data are used.
The first dataset is introduced in~\citep{Lluis2020} and consists of 5,000 rectangular rooms.
The room dimensions are sampled randomly in accordance with the recommendations for listening rooms in ITU-R BS.1116-3 \citep{BS1116}.
The dataset uses a constant reverberation time $T_{60}$ of 0.6 s and only includes room modes in the $x$ and $y$ dimensions, i.e. $n_z=0$.

The second dataset consists of 20,000 rectangular rooms.
Room dimensions are uniformly sampled with $V\sim \mathcal{U}(50,300)\text{m}^3$, $l_x\sim \mathcal{U}(3.5,10)$m, $l_z\sim \mathcal{U}(1.5,3.5)$m, and $l_y=V/l_xl_z$.
Compared to the first dataset, the room dimensions span a larger range and allow us to represent e.g. the Product Room, which is not included in the original training data. 
The dataset uses reverberation times $T_{60}$ sampled from $\mathcal{U}(0.2, 1.0)$s and includes room modes in all three $x$-, $y$-, and $z$-dimensions.

For both datasets, a grid $\mathcal{D}_o^{L,P}$ is defined with $I=J=8$ and $L=P=4$, which effectively divides a sound field into 32x32 uniformly-spaced microphone positions.
Using this grid, the magnitude of the sound field is reconstructed at 1/12 octave center-frequencies resolution in the range [30, 300]~Hz. Simulations are specified to include all room modes with a resonance frequency below 400~Hz, which means that there is a total of $K=40$ frequency slices.

\subsection{Experiments on the ISOBEL Sound Field Dataset}\label{sec:mag_isobel_exp}
\begin{figure}[tb]
\centering
\includegraphics[width=\linewidth]{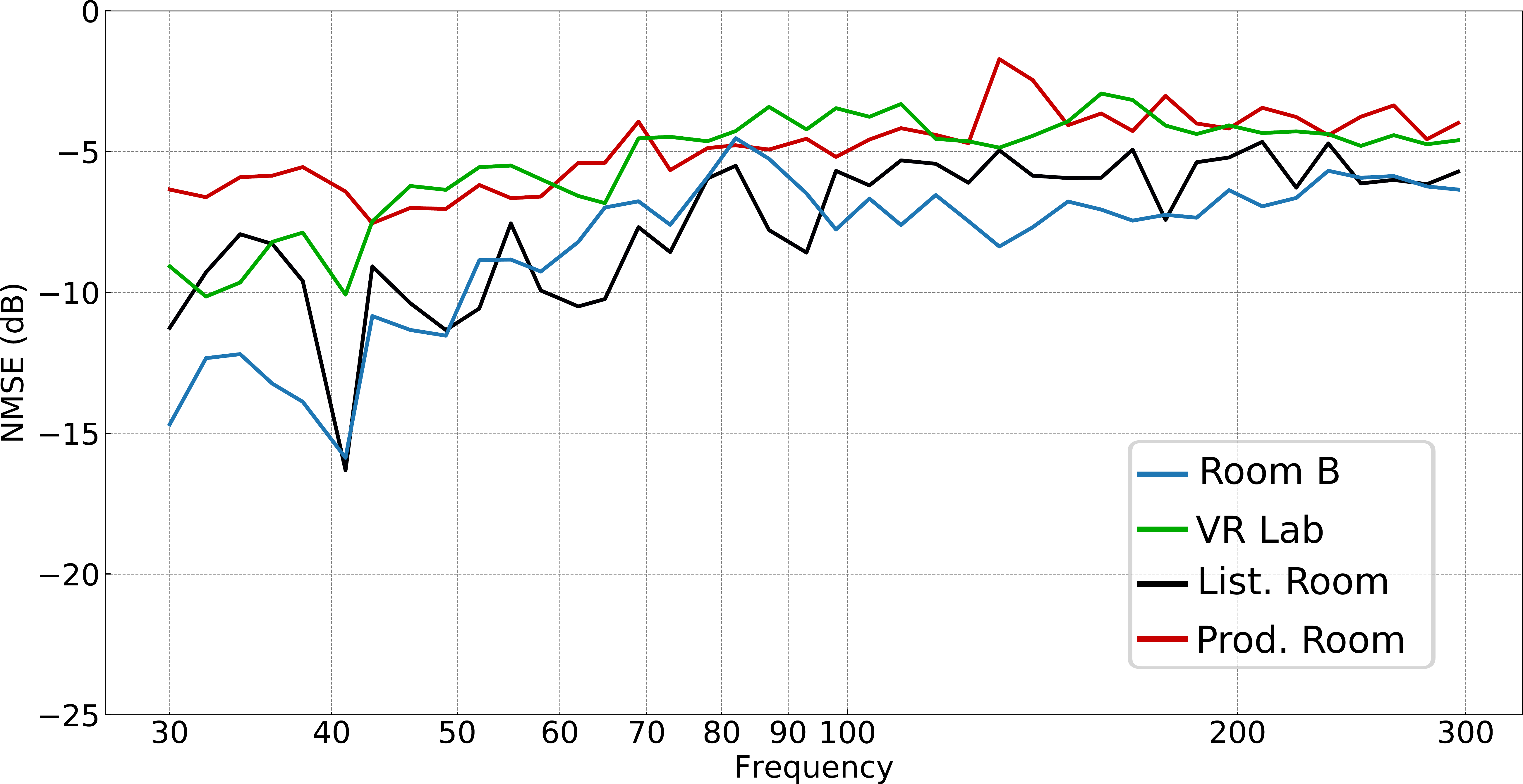}
\caption{NMSE in dB of U-Net-based magnitude reconstruction in the four measured rooms with $n_{mic}=15$ using the original pretrained model presented in~\citep{Lluis2020}.}
\label{fig:mag_res1}
\end{figure}
\begin{figure}[tb]
\centering
\includegraphics[width=\linewidth]{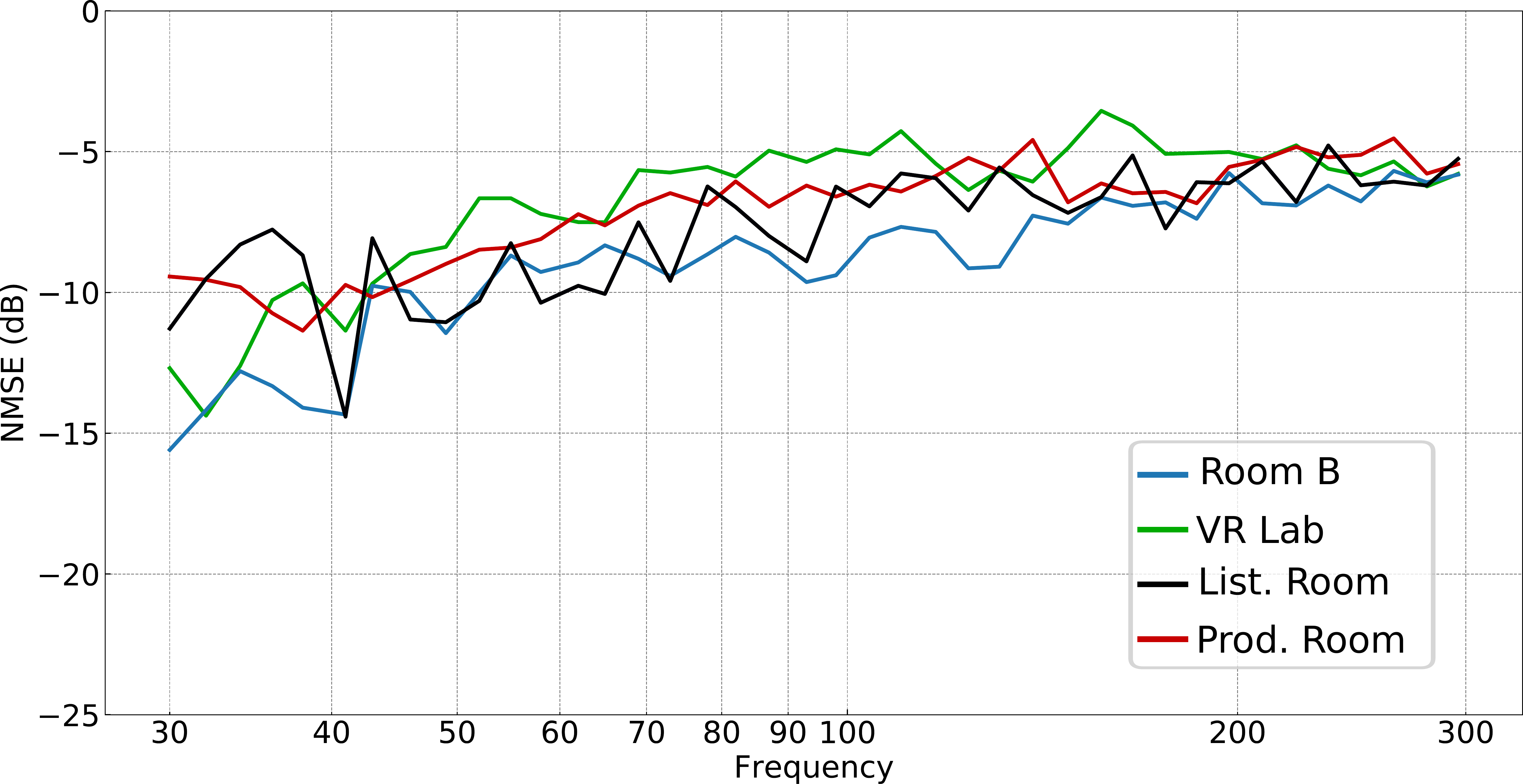}
\caption{NMSE in dB of U-Net-based magnitude reconstruction in the four measured rooms with $n_{mic}=15$ using the model presented in~\citep{Lluis2020} trained using the extended dataset.}
\label{fig:mag_res2}
\end{figure}
The U-Net-like architecture has shown promising results on simulated data and on measurements from a single real room \citep{Lluis2020}.
In the following experiments, we expose the model to the ISOBEL Sound Field dataset.
We include results from the original model, as well as a model built around a similar architecture but using the extended training data with a larger range of room dimensions and reverberation characteristics.
We investigate the performance of the model trained with the two different simulated datasets in the four rooms included in the real-world dataset.
Special attention is paid to the number of available samples, i.e. the number of microphones $n_{mic}$.
We are mainly interested in settings with a very low number of microphones.
In particular, we show results for 5, 15, and 25 microphones in the rooms with a total of $32\times 32=1024$ available positions.
In each room, a total of 40 different and randomly sampled realizations of microphone positions $\mathcal{S}_o$ are used for each value of $n_{mic}$.
We report the average performance across the 40 realizations, and use the source located in one of the corners of each room.

\begin{table}[b]
\caption{MNMSE in dB with $M=40$ different and randomly sampled realizations of $\mathcal{S}_o$ for each room in the ISOBEL SF dataset. A lower score is better.}
\label{tab:mag_mnmse}
\newcolumntype{P}[1]{>{\centering\arraybackslash}p{#1}}
    \begin{tabular}{p{2.0cm}P{1.5cm}P{1cm}P{1cm}P{1cm}}
\toprule
    &&\multicolumn{3}{c}{$n_{mic}$}\\\cline{3-5}
    Room & Model & 5 & 15 & 25\\\midrule
    \multirow{2}{*}{Room B}     & Orig. & -6.33 & -8.71 & -9.62\\
                                & Ext.  & -6.27 & -8.84 & -10.25\\\cline{2-5}
    \multirow{2}{*}{VR Lab}     & Orig. & -4.01 & -5.08 & -5.63\\
                                & Ext.  & -4.12 & -6.78 & -8.05\\\cline{2-5}
    \multirow{2}{*}{List. Room} & Orig. & -4.38 & -6.92 & -7.94\\
                                & Ext.  & -5.00 & -7.61 & -8.44\\\cline{2-5}
    \multirow{2}{*}{Prod. Room} & Orig. & -3.89 & -4.91 & -5.55\\
                                & Ext.  & -5.18 & -6.67 & -7.73\\
\bottomrule
\end{tabular}
\end{table}

\begin{figure*}[tb]
    \centering
    \includegraphics[width=\textwidth]{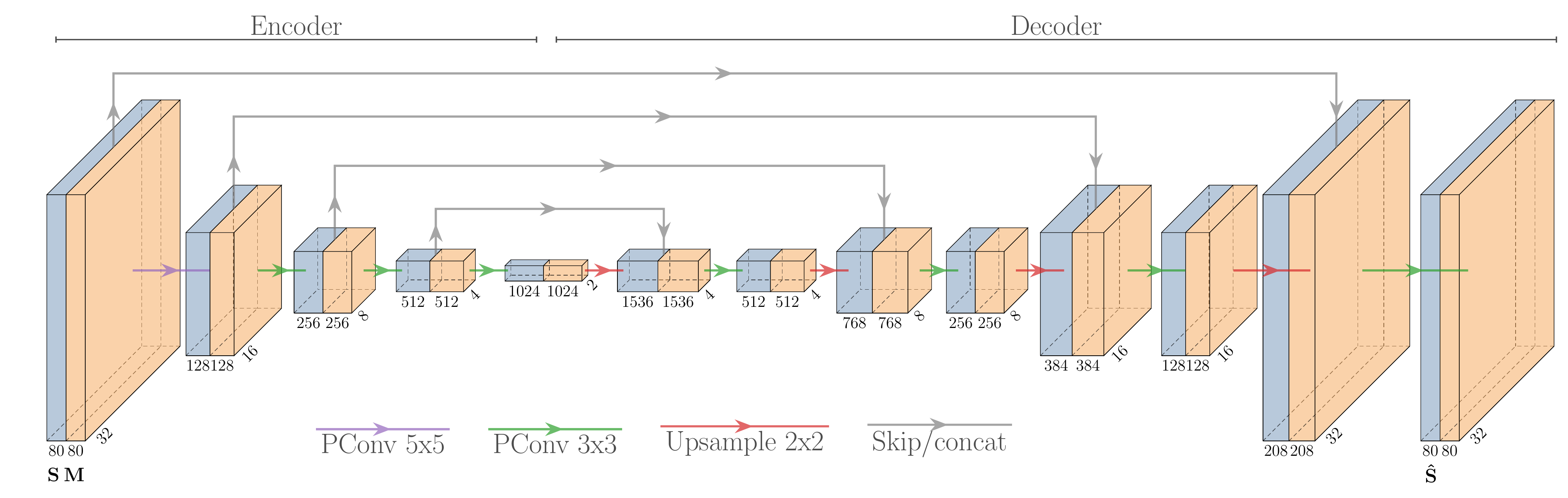}
    \caption{Architecture of the U-Net-like convolutional neural network proposed for complex sound field reconstruction. $\mathbf{S}$ is the tensor with real and imaginary sound fields concatenated along the frequency-dimension, $\mathbf{M}$ is the mask tensor, and $\mathbf{\hat{S}}$ is the reconstructed sound field tensor.}
\label{fig:unet_complex}
\end{figure*}

Fig.~\ref{fig:mag_res1} and Fig.~\ref{fig:mag_res2} show NMSE$_k$ results for 15 microphones of model trained with the original and the extended datasets, respectively.
It is clear that the model trained with the original dataset does not generalize well to all the rooms.
This behavior is expected, since the training data are not designed to represent rooms that fall outside the recommendations for listening room dimensions.
On the contrary, the extended training data are motivated in encompassing a wider selection of rooms, which also shows in the results for e.g. the Product Room.
One important observation in this regard is that performance does not decrease in rooms that are already represented in the simulated data when more diverse simulated rooms are included, which can e.g. be seen from the performance in Room B.
This result indicates that the capacity of the model is sufficient for generalizing to a wide range of diverse rooms and room acoustic characteristics, given that the model is provided with ample training samples.

Table~\ref{tab:mag_mnmse} details MNMSE results, which are the NMSE results averaged across frequencies $K=40$ and $\mathcal{S}_o$ realizations $M=40$.
The MNMSE results for $n_{mic}=15$ are the condensed results shown for the NMSE$_k$ in Figs.~\ref{fig:mag_res1} and \ref{fig:mag_res2}.
The scores in the table reiterate the observations from the figures, performance is improved with the extended training data for some rooms in particular, while performance is maintained in the other rooms.
Interestingly, there seems to be a tendency of more pronounced improvements with a larger number of microphones.
We attribute this effect to similar observations within classical methods that as the number of microphones increase, relative improvement for reconstruction is higher at low frequencies as opposed to the high-frequency range, \citep{Lluis2020,Ajdler2006}.

In summary, the deep learning-based model is confirmed to possess the ability to generalize to a diverse set of real rooms for sound field magnitude reconstruction.
Based solely on training with simulated data, these promising results motivate further investigations, e.g. of reconstructing the complex-valued sound fields.

\section{Complex Sound Field Reconstruction}\label{sec:complex}

We propose to extend the U-Net-based model to work with complex-valued room transfer functions (RTFs).
Reconstruction of both magnitude and phase of sound fields enable new opportunities, such as the application of sound zones.
A topic, which we investigate in Section~\ref{sec:sound_zones}.

The proposed model is based on the model designed to work with the magnitude of sound fields.
Note that deep learning-based models that work directly on complex-valued inputs have been introduced, e.g. within Transformers \citep{Yang2020,Kim2020b}, but in this paper we instead choose to process the sound fields such that the U-Net-based model receives real-valued inputs.
Specifically, we present the model to real and imaginary parts of sound fields separately.
That is, where the magnitude-based model receive as input $\{|s(\mathbf{r},\omega_k)|\}_{\mathbf{r}\in\mathcal{D}_o^{L,P},k}$ in the tensor form $\mathbf{S}_{mag}\in\mathbb{R}^{IL \times JP \times K}$, the complex-based model instead receives a concatenation of the real and imaginary sound fields.
Specifically, using the real sound field $\{s_\text{Re}(\mathbf{r},\omega_k)\}_{\mathbf{r}\in\mathcal{D}_o^{L,P},k}$ with the tensor form $\mathbf{S}_\text{Re}\in\mathbb{R}^{IL \times JP \times K}$, and similarly the imaginary sound field tensor $\mathbf{S}_\text{Im}\in\mathbb{R}^{IL \times JP \times K}$, we define the concatenated input:
\begin{equation}
    \mathbf{S}:=\left[\mathbf{S}_\text{Re}\ \mathbf{S}_\text{Im}\right],
\end{equation}
where $\mathbf{S}\in\mathbb{R}^{IL \times JP \times 2K}$ is the resulting tensor with real and imaginary sound fields concatenated along the frequency-dimension.
Note that the complex-valued sound field is easily recovered from this tensor form.
In addition, we define a mask tensor $\mathbf{M}\in\mathbb{R}^{IL \times JP \times 2K}$ computed from $\mathcal{S}_o$ and $\mathcal{D}_o^{L,P}$.

We follow the pre- and postprocessing steps as described in~\citep{Lluis2020}, which entails completion, scaling, upsampling, mask generation, and rescaling based on linear regression.
These steps are, however, adjusted such that they operate on a tensor that has doubled in size from $K$ to $2K$ in the third dimension.
Furthermore, we have observed significant improvements by changing the min-max scaling of the input to a max scaling that takes into account both real and imaginary parts for each frequency slice.
Specifically:
\begin{equation}
    s_{\text{Re},s}(\mathbf{r},\omega_k):=\frac{s_\text{Re}(\mathbf{r},\omega_k)}{\max_{\mathbf{r}\in\mathcal{S}_o}\left(|s_\text{Re}(\mathbf{r},\omega_k)|,|s_\text{Im}(\mathbf{r},\omega_k)|\right)}
\end{equation}
\begin{equation}
    s_{\text{Im},s}(\mathbf{r},\omega_k):=\frac{s_\text{Im}(\mathbf{r},\omega_k)}{\max_{\mathbf{r}\in\mathcal{S}_o}\left(|s_\text{Re}(\mathbf{r},\omega_k)|, |s_\text{Im}(\mathbf{r},\omega_k)|\right)}
\end{equation}
for each $\omega_k$.
Note that this alters the scaling operation from working in the range [0,1] to working in [-1,1].
The motivation in doing so, is that values can be negative, in contrast to the real values from the magnitude.
By using max scaling we ensure that zero will not shift between realizations.

The architecture of the proposed neural network, as illustrated in Fig.~\ref{fig:unet_complex}, is based on a U-Net \citep{Ronneberger2015}.
We employ partial convolutions (PConv) as proposed for image inpainting in~\citep{Liu2018}.
In the encoding part of the U-Net, we use a stride of two in the partial convolutions in order to halve the feature maps, while doubling the number of kernels in each layer.
The decoder part acts opposite with upsampling feature maps and reducing the number of kernels to reach an output tensor $\mathbf{\hat{S}}$ with matching dimensions to the input tensor $\mathbf{S}$.
We use ReLU as activation function in the encoding part, and leaky ReLU with a slope coefficient of -0.2 in the decoder.
We initialize the weights using the uniform Xavier method \citep{Glorot2010}, initialize the biases as zero, and use the Adam optimizer \citep{Kingma2014} with early stopping when performance on a validation set stops increasing.
Due to the increased input and output sizes, we double the number of kernels in all layers compared to the U-Net for magnitude reconstruction.
We also do not use a 1x1 convolution with sigmoid activation in the last layer, since the range of our output is not constrained to [0,1] but instead [-1,1].
We have not experienced any decreases in performance from not including this layer.

\subsection{Experiments}
In this section, we assess the complex-valued sound field reconstruction.
The simulated extended dataset introduced in Section~\ref{sec:sim_data} is used to train the model.
It is important to note that NMSE scores are not directly comparable between magnitude and complex reconstruction, for which reason it is not possible to scrutinize differences between the two types of models.
That is, the results presented in the following experiments will stand on their own, and only indicative parallels can be drawn to the results from magnitude reconstruction.

\begin{figure} [tb]
    \centering
    \includegraphics[width=\linewidth]{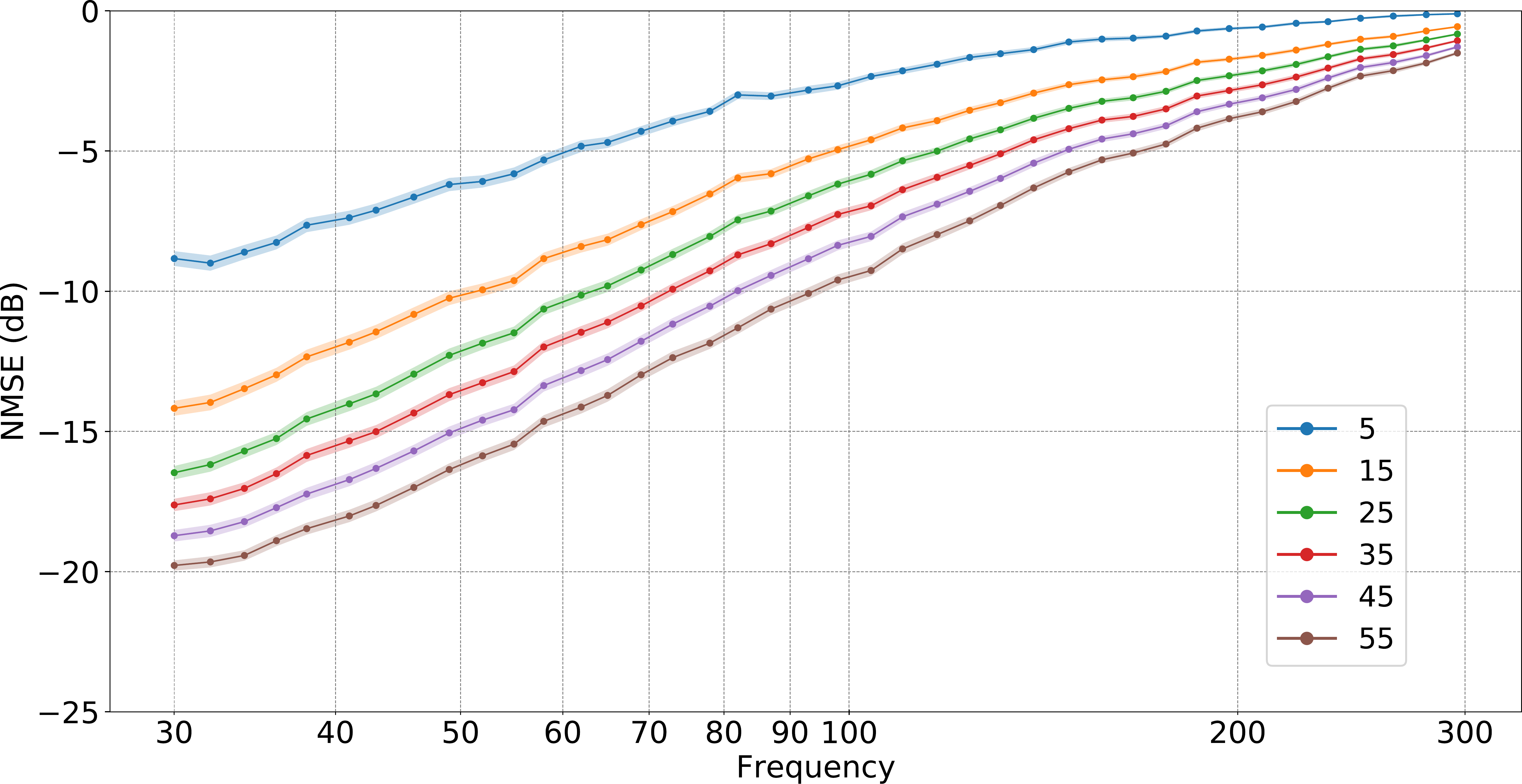}
    \caption{NMSE in dB for complex reconstruction of simulated sound fields in the test set with 190 different rooms and three realizations of $\mathcal{S}_o$ in each room ($M=570$ for each value of $n_{mic}$). The solid lines indicate average NMSE$_k$ shown with 95\% confidence intervals. Colors indicate different values of $n_{mic}$ in the range [5, 55].}
\label{fig:complex_sim}
\end{figure}

First, we test how the model performs on the simulated data associated with the training data, but held out specifically for evaluation.
This test set consists of 190 simulated rooms, the validation set contains approximately 1,000 rooms, and the training set holds the remaining rooms from the 20,000 available rooms. 
In each room, three different realizations of $\mathcal{S}_o$ are used for each value of $n_{mic}$.
Results in terms of NMSE are shown in Fig.~\ref{fig:complex_sim}. 
Some tendencies are similar to those observed for magnitude reconstruction, such as improvements in performance with an increasing number of available microphones. 
At the same time, as frequency increases, performance degrades.

\begin{figure}[tb]
\centering
\includegraphics[width=\linewidth]{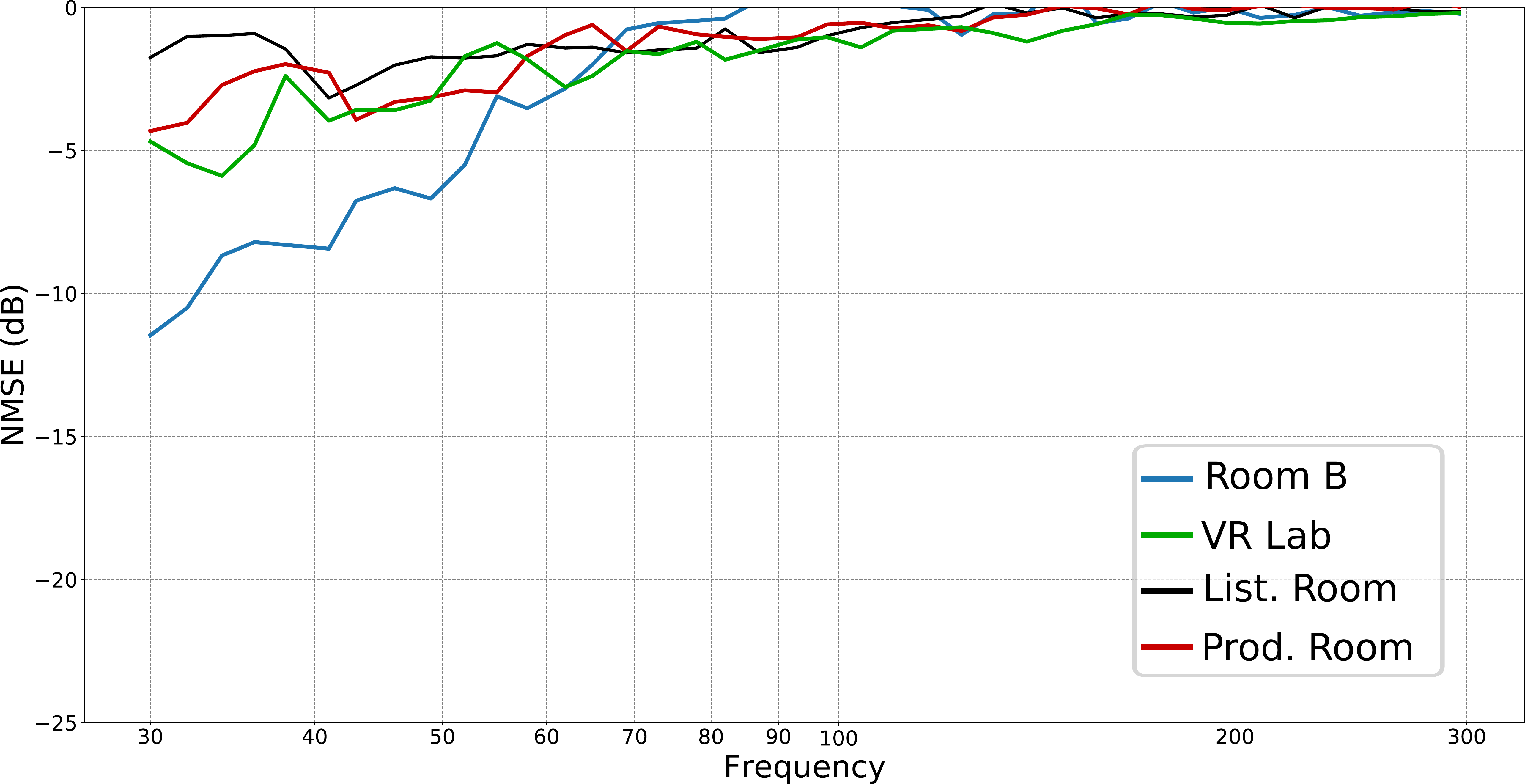}
\caption{Average NMSE$_k$ in dB of complex reconstruction in the four measured rooms with $n_{mic}=15$.}
\label{fig:complex_res}
\end{figure}

Next, we evaluate the complex reconstruction model on the ISOBEL Sound Field dataset.
The approach is similar to the experiment in Section~\ref{sec:mag_isobel_exp}, except the use of the complex-valued sound fields instead of the magnitude.
As can be seen from the results in Fig.~\ref{fig:complex_res}, performances in the real rooms are not comparable to those from simulated data.
Moreover, although it is not possible to compare directly, performance seems worse than what is achieved with the magnitude-based reconstruction in the same rooms, see Fig.~\ref{fig:mag_res2}.
That is, the complex reconstruction model is not transferring useful knowledge as successfully  from the simulations-based training to the real world.
Given that the network is able to reconstruct the simulated sound fields, it appears that the complex simulation model is a worse match for the real rooms than the magnitude simulation model. 
The outcome is that the framework is able to reconstruct sound fields which are close to fields included in the training data, it is indicated that the complex simulations are a poor match for the real rooms. 
Two apparent differences are the identical boundary conditions at all surfaces and perfectly rectangular geometry assumed in the simulations, but which are not true in the real rooms. 
To provide insights into how the network behaves relative to rooms which does not match the training data set we now present the following simulations.


\begin{figure*}[tb]
    \begin{minipage}[c]{0.2\linewidth}%
        \centering
        \tikz{\node[below left, inner sep=3pt] (Train) {Train$\downarrow$};%
      \node[above right,inner sep=3pt] (Test) {Test$\rightarrow$};%
      \draw (Train.north west|-Test.north west) -- (Train.south east-|Test.south east);}
    \end{minipage}%
    \begin{minipage}[c]{0.25\linewidth}%
        \centering Simulated\\ List. Room\\ \phantom{.}
    \end{minipage}%
    \begin{minipage}[c]{0.25\linewidth}%
        \centering Simulated\\ List. Room\\ $l_x + \mathcal{U}(-0.25,0.25)$m
    \end{minipage}%
    \begin{minipage}[c]{0.25\linewidth}%
        \centering Simulated\\ List. Room\\ $l_x + \mathcal{U}(-1.0,1.0)$m
    \end{minipage}\\\vspace{1em}
    \begin{minipage}[c]{0.2\linewidth}%
    \centering%
    Simulated\\ List. Room
    \end{minipage}%
    \begin{minipage}[c]{0.25\linewidth}%
        \centering
        \includegraphics[width=0.95\linewidth]{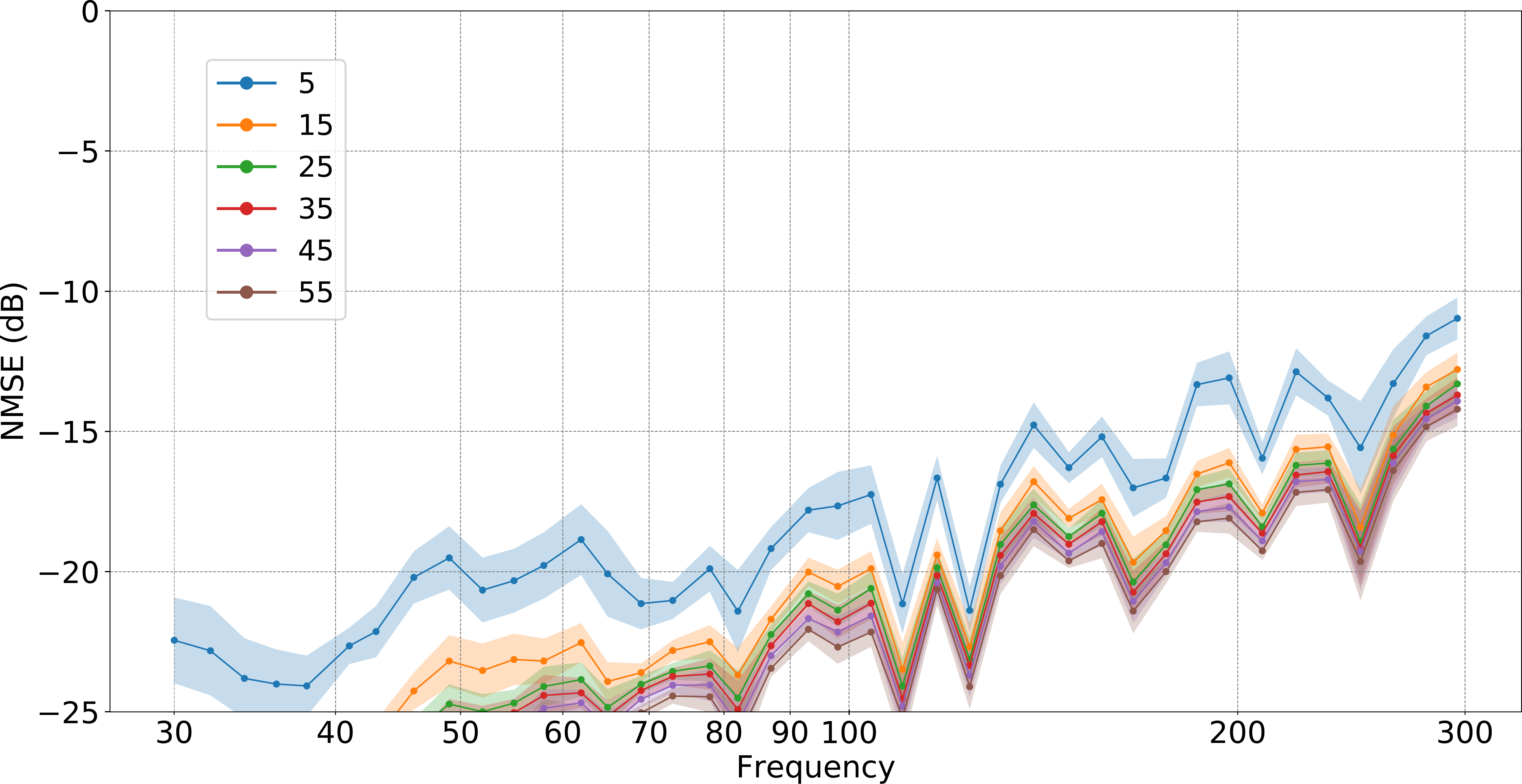}
    \end{minipage}%
    \begin{minipage}[c]{0.25\linewidth}%
        \centering
        \includegraphics[width=0.95\linewidth]{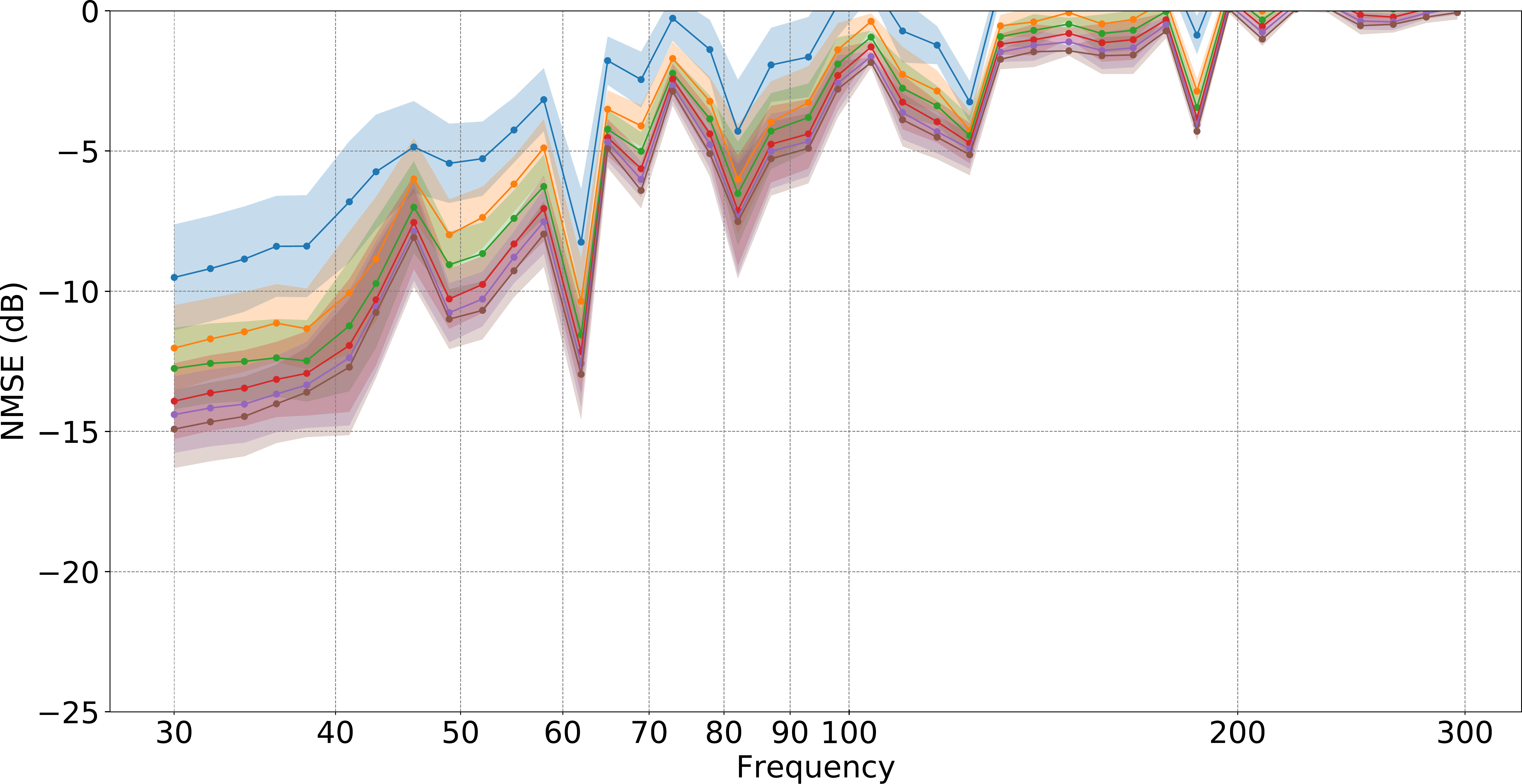}
    \end{minipage}%
    \begin{minipage}[c]{0.25\linewidth}%
        \centering
        \includegraphics[width=0.95\linewidth]{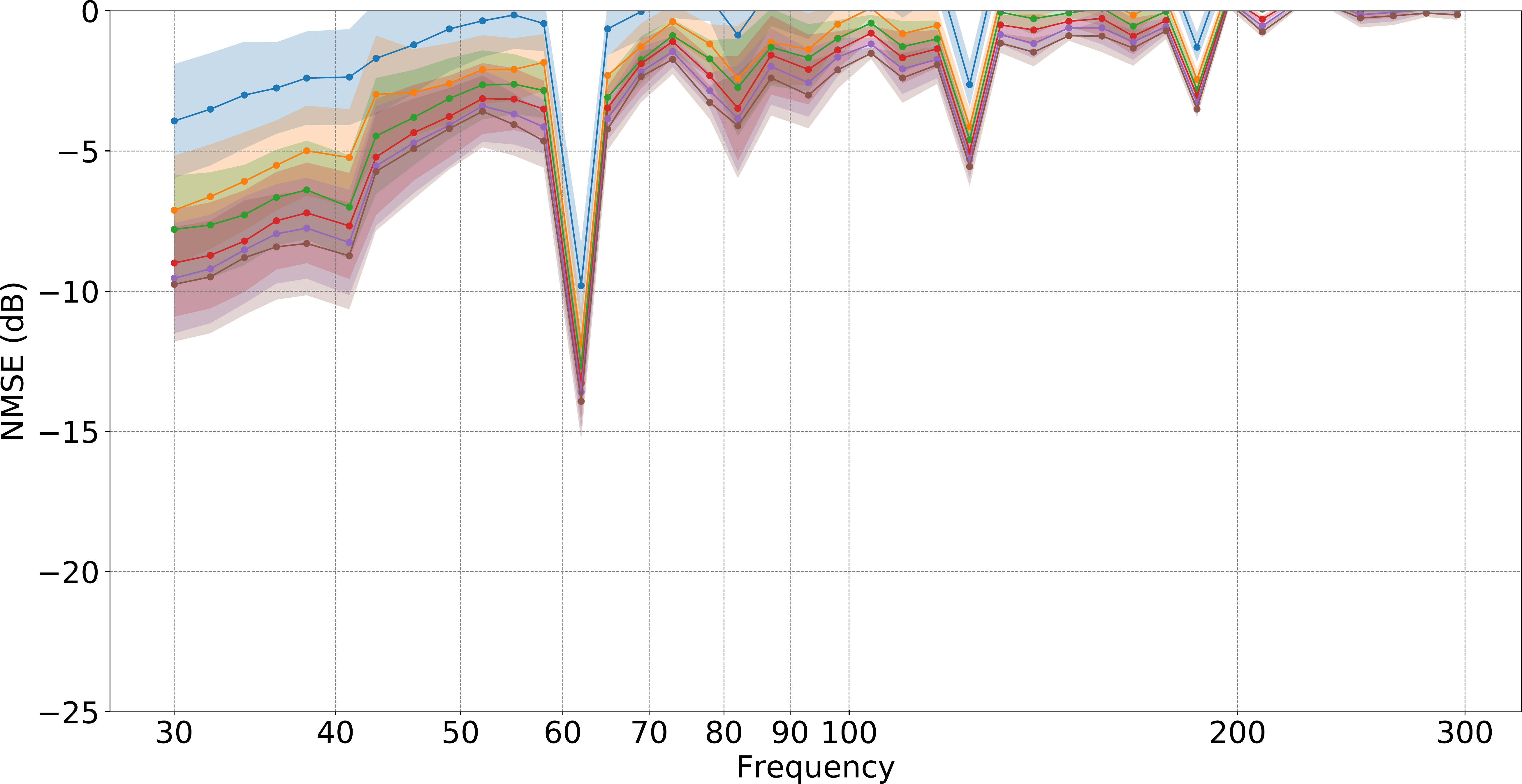}
    \end{minipage}\\\vspace{1em}
    \begin{minipage}[c]{0.2\linewidth}%
        \centering Simulated\\ List. Room\\ $l_x + \mathcal{U}(-0.25,0.25)$m
    \end{minipage}%
    \begin{minipage}[c]{0.25\linewidth}%
        \centering
        \includegraphics[width=0.95\linewidth]{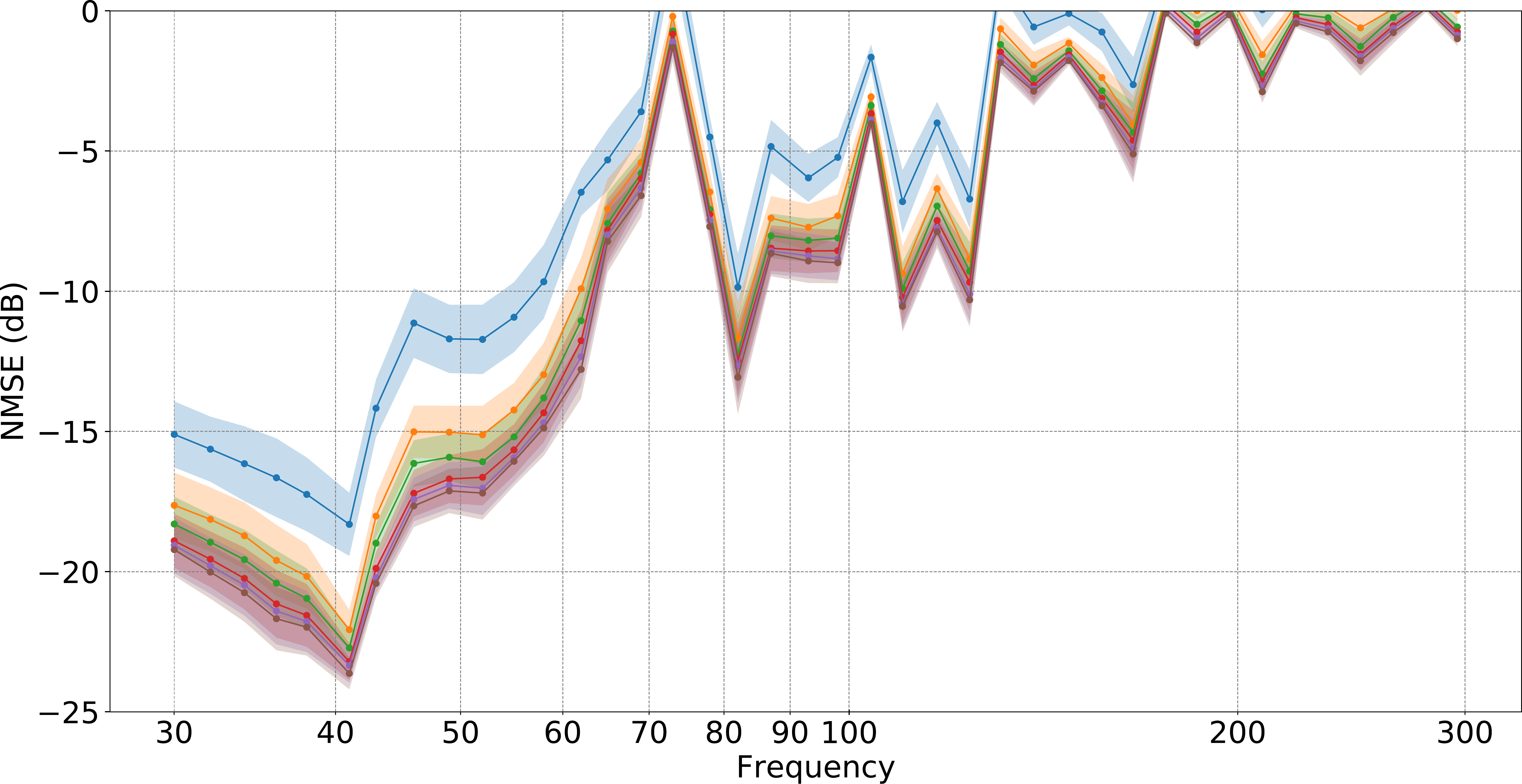}
    \end{minipage}%
    \begin{minipage}[c]{0.25\linewidth}%
        \centering
        \includegraphics[width=0.95\linewidth]{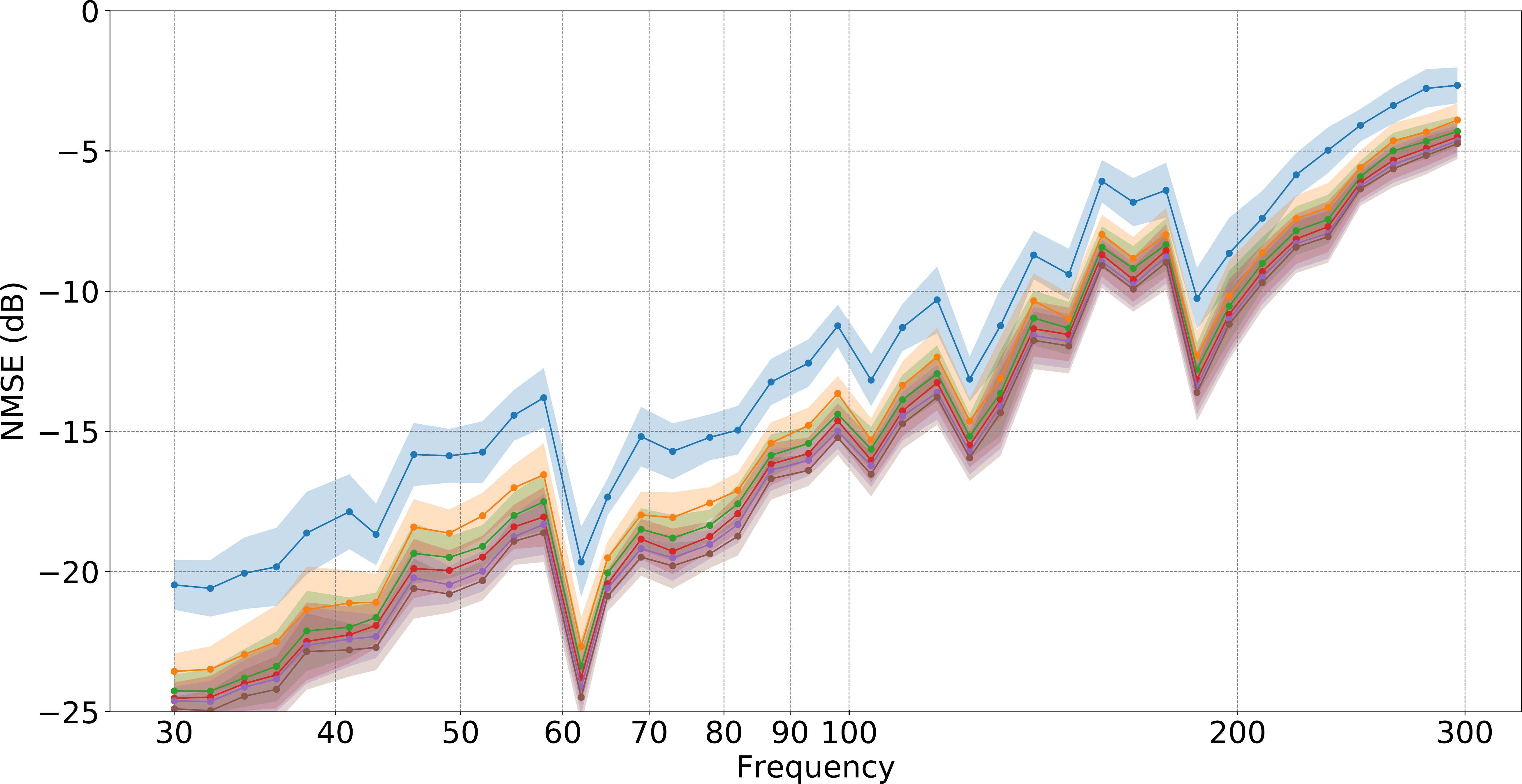}
    \end{minipage}%
    \begin{minipage}[c]{0.25\linewidth}%
        \centering
        \includegraphics[width=0.95\linewidth]{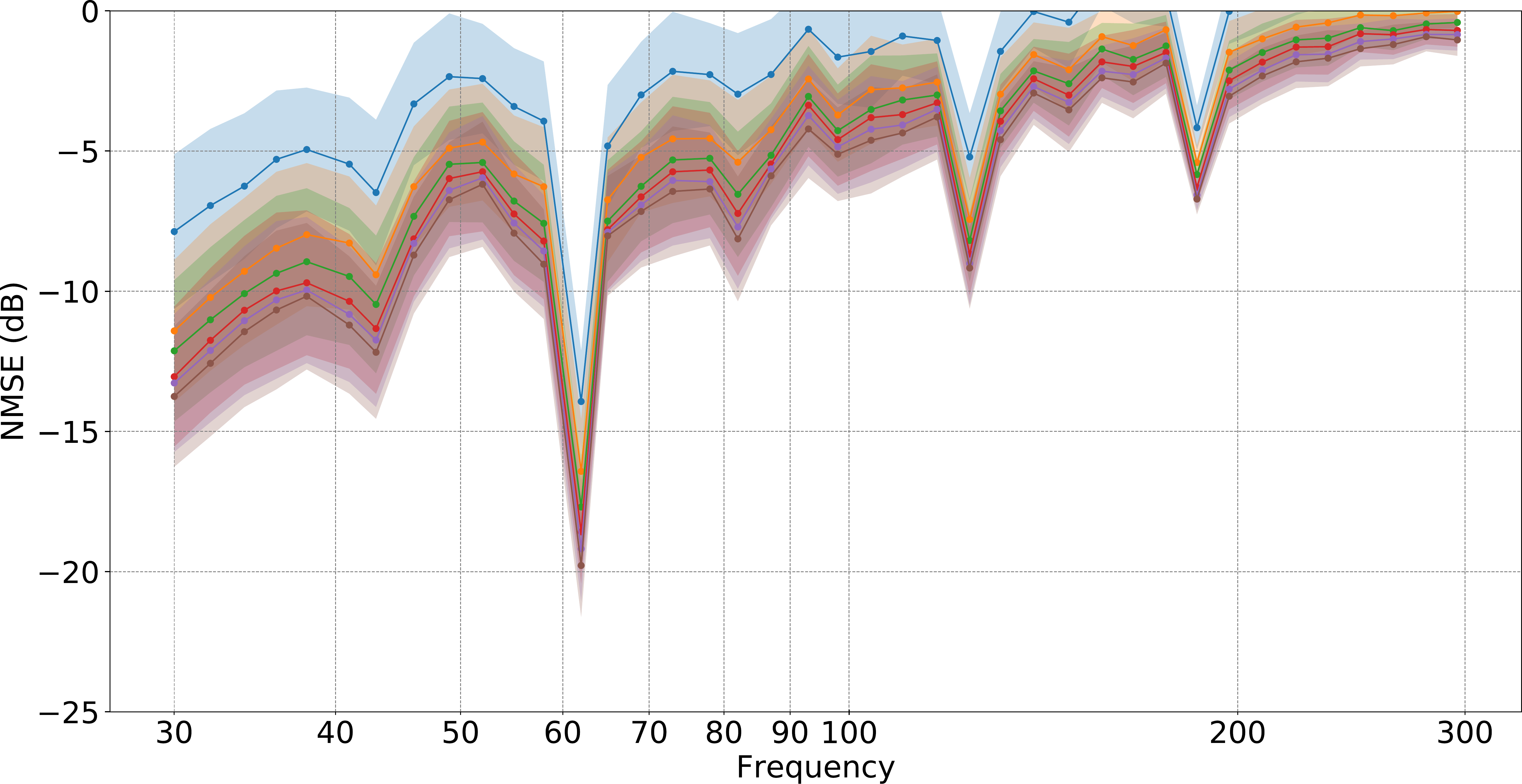}
    \end{minipage}\\\vspace{1em}
    \begin{minipage}[c]{0.2\linewidth}%
        \centering Simulated\\ List. Room\\ $l_x + \mathcal{U}(-1.0,1.0)$m
    \end{minipage}%
    \begin{minipage}[c]{0.25\linewidth}%
        \centering
        \includegraphics[width=0.95\linewidth]{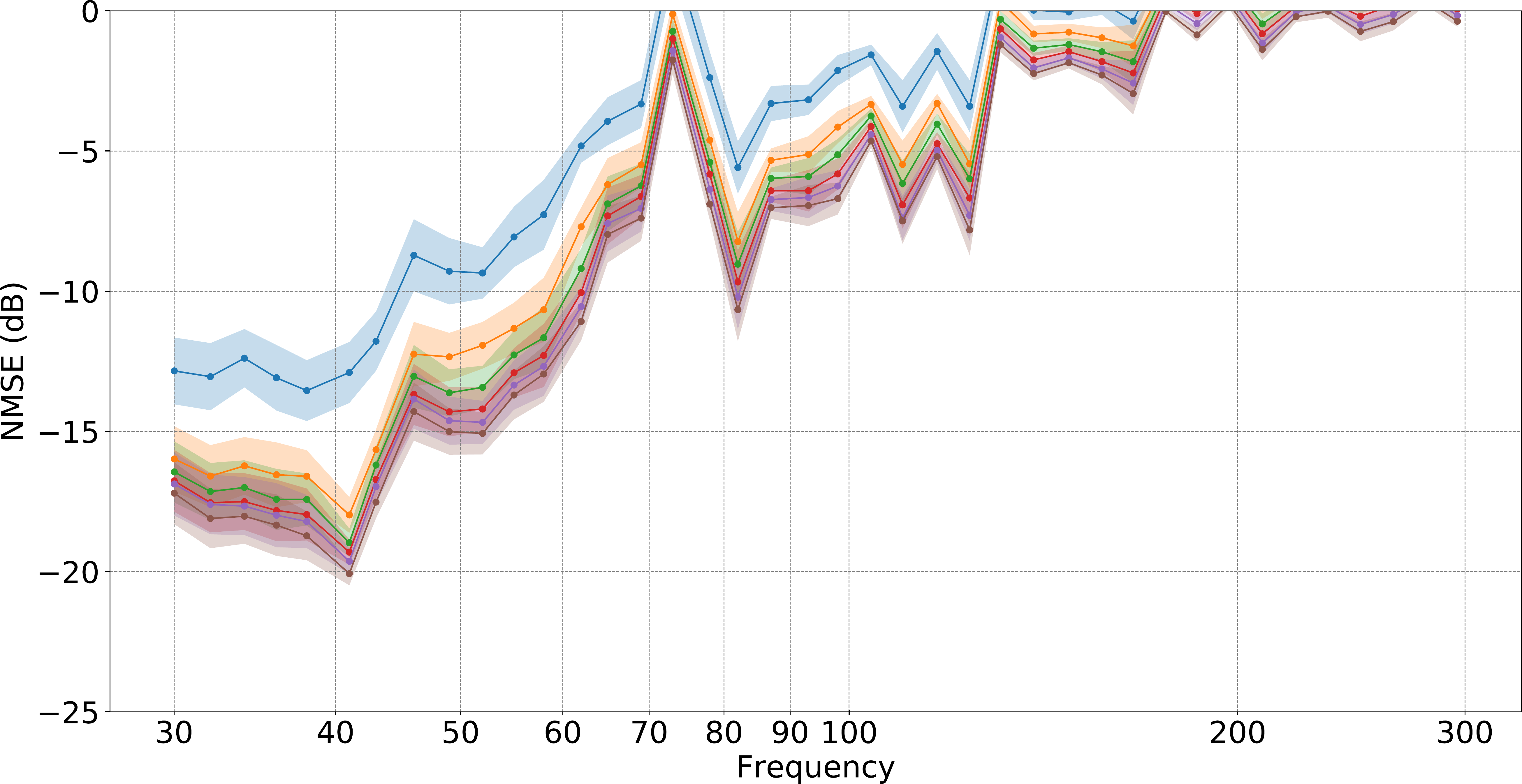}
    \end{minipage}%
    \begin{minipage}[c]{0.25\linewidth}%
        \centering
        \includegraphics[width=0.95\linewidth]{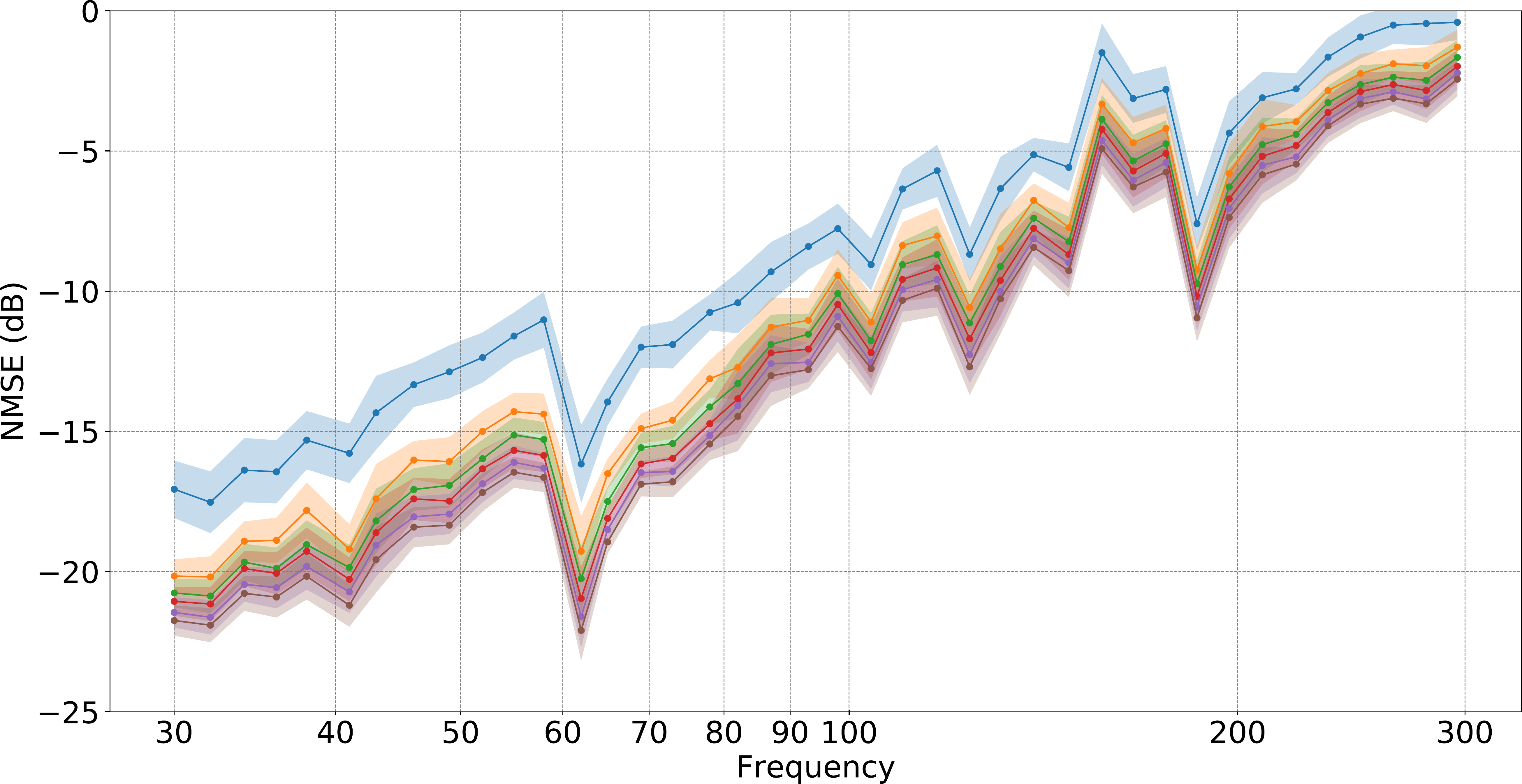}
    \end{minipage}%
    \begin{minipage}[c]{0.25\linewidth}%
        \centering
        \includegraphics[width=0.95\linewidth]{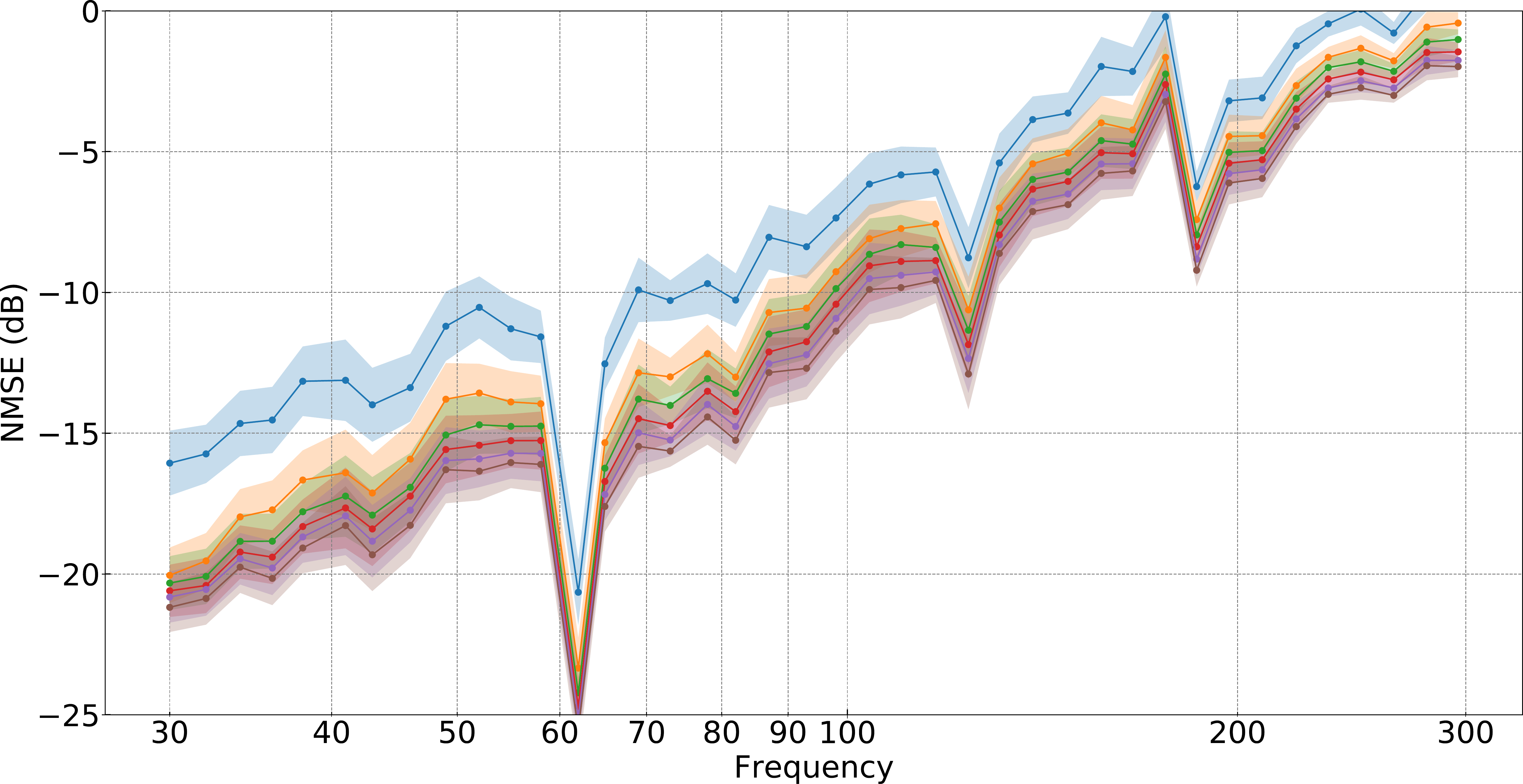}
    \end{minipage}%
    \caption{NMSE in dB for complex reconstruction of simulated sound fields in rooms with no or small variations in the room dimensions. Rows: Training data. Columns: Test data. Four random realizations of $\mathcal{S}_o$ are used in each of the 11 test rooms ($M=44$). The solid lines indicate average NMSE$_k$ shown with 95\% confidence intervals. Colors indicate different $n_{mic}$ values, i.e., $n_{mic}=5$ (blue), $n_{mic}=15$ (orange), $n_{mic}=25$ (green), $n_{mic}=35$ (red), $n_{mic}=45$ (purple), and $n_{mic}=55$ (brown).}
\label{fig:complex_list_room}
\end{figure*}

\subsection{Discussion of Experiments}
Several optimizations and fine-tuning approaches have been investigated for the complex reconstruction in real rooms without achieving notable improvements.
Instead, we take another approach, and show what happens to the model, when it is exposed to data that are not represented in the training data.
To this end, we are interested in assessing the performance of room specialized models.
That is, if room dimensions and reverberation time are known, how well will a model trained specifically for that room perform.
For this, we introduce new datasets each with 824 realizations for training, 165 for validation, and 11 for testing.
Each simulated realization has a randomly positioned source.
In total, three such datasets are generated according to the procedure described in Section~\ref{sec:sim_data}.
The first dataset assumes that room characteristics are known perfectly, we use the parameters of the Listening Room.
The second and third datasets introduce uncertainty in the room dimensions.
In particular, we alter the length and width of rooms, while keeping the aspect ratio ($l_x/l_y$) of the room constant.
We accomplish this by uniformly sampling an error, which is added to the length of a room, and correct the width to achieve the original aspect ratio.
The two datasets sample errors in the range [-0.25, 0.25] m and [-1, 1] m, respectively.
The results for the three models evaluated on each of the test sets are shown in Fig.~\ref{fig:complex_list_room}.
The first column shows how the three models perform on the dataset with no added uncertainties.
Even with small variations of the 0.25 m scale, performance rapidly degrades with increasing frequency.
On the diagonal, training data match test data, and once again high frequencies see a significant performance decrease with increasing uncertainty.
In general, the models do not perform well on datasets with more variation than what is included in their own training data, which can be seen in the three upper right figures.

Further experiments showed that the three models do not generalize to the real-world measurements of the Listening Room.
This result indicates that the simplifications imposed during the simulations of rooms causes the simulated sound fields to not represent the exact real rooms we intend it to.
That is, a model trained with simulated data generated using exact parameters of a real room will not be able to reconstruct the sound field accurately in the real room.
As suggested by our results, neither will a model trained with $\pm1$ m uncertainty.
This calls for inclusion of diverse room parameters when training a model with simulated data if the intended purpose is to use the reconstruction in real rooms.

We showed in Section~\ref{sec:mag} how magnitude reconstruction recovered performance in some of the real rooms by using an extended training dataset with more diverse simulated rooms.
The same effect is not observed for complex reconstruction.
We believe two factors are the main reasons: 1) the boundary conditions in the simulations assume nearly rigid walls and do not include e.g. phase shifts of real wall reflections; 2) the simulations assume perfectly rectangular rooms with a uniform distribution of absorption.
Thus, we hypothesize that the model does not see representative data during training, analogous to not having the correct room dimensions represented in the training data.
\section{The Sound Zones Application}\label{sec:sound_zones}
One potential application for the sound field reconstruction presented in this paper, is in the process of setting up sound zones. Sound zones generally refers to the scenario where multiple loudspeakers are used to reproduce individual audio signals to individual people within a room \citep{Betlehem2015}. To control the sound field at the location of the listeners in the room, it is necessary to know the RTFs between each loudspeaker and locations sampling the listening regions. If the desired locations of the sound zones change over time, it becomes labor intensive to measure all the RTFs in situ. As an alternative, a small set of RTFs could be measured and used to extrapolate the RTFs at the positions of interest.

\subsubsection{Setup}
For this example, we will explore the scenario where sound is reproduced in one zone (the bright zone) and suppressed in another zone (the dark zone).\footnote{The use case with multiple individual audio signals can be realized using superposition of this solution and one where the role of bright and dark zone are reversed.}

The question posed in a sound zones scenario, is how the output of the available loudspeakers should be adjusted to achieve the desired scenario. A simple formulation of this problem in the frequency domain is typically denoted acoustic contrast control and relies on maximizing the ratio of mean square pressure in the bright zone relative to the dark zone \citep{Choi2002}. This ratio is termed as the acoustic contrast and can be expressed as
\begin{equation}
  \text{Contrast}(\omega) := \frac{\|\mathbf H_B(\omega) \mathbf q(\omega)\|_2^2}{\|\mathbf H_D(\omega) \mathbf q(\omega)\|_2^2}
\end{equation}
where $\mathbf H_B(\omega) \in \mathbb C^{M\times L}$ is a matrix of RTFs from $L$ loudspeakers to $M$ microphone positions in the bright zone and $\mathbf H_D(\omega) \in \mathbb C^{M\times L}$ are the RTFs from the loudspeakers to points in the dark zone.  The adjustment of the loudspeaker responses $\mathbf q(\omega) \in \mathbb C^L$ can be determined as the eigenvector of $(\mathbf H_D^H(\omega) \mathbf H_D(\omega) + \lambda_D \mathbf I)^{-1}\mathbf H_B^H(\omega) \mathbf H_B(\omega)$ which corresponds to the maximal eigenvalue \citep{Elliott2012}, where $\cdot^H$ denotes the Hermitian transpose. In this investigation, the regularization parameter is chosen as
\begin{equation}
  \lambda_D = 0.01 \|\mathbf H_D^H(\omega) \mathbf H_D(\omega)\|_2.
\end{equation}
This choice is made to scale the regularization relative to the maximal singular value of $\mathbf H_D^H(\omega) \mathbf H_D(\omega)$, thereby, controlling the condition number of the inverted matrix.

\subsubsection{Sparse Reconstruction method}
An alternative method for estimating the RTFs at positions of interest can be obtained by a sparse reconstruction problem inspired by \citep{Fernandez-grande2019}. Here, the sound pressure observed at the physical microphone locations are modeled as a combination of impinging plane waves
\begin{equation}
  \underset{\mathbf s(\omega)}{\underbrace{
    \begin{bmatrix}
      s(\mathbf r_1,\omega)\\
      \vdots\\
      s(\mathbf r_M,\omega)
    \end{bmatrix}
  }} =
  \underset{\boldsymbol \Phi}{\underbrace{
    \begin{bmatrix}
      \phi_1(\mathbf r_1) & \cdots & \phi_N(\mathbf r_1)\\
      \vdots & \ddots & \vdots\\
      \phi_1(\mathbf r_M) & \cdots & \phi_N(\mathbf r_M)
    \end{bmatrix}
  }}
  \underset{\mathbf b(\omega)}{\underbrace{
    \begin{bmatrix}
      b_1(\omega)\\
      \vdots\\
      b_N(\omega)
    \end{bmatrix}
  }}
\end{equation}
where $\mathbf s(\cdot,\cdot)$ is defined in (\ref{eq:2}), $\phi_n(\mathbf r_m) = e^{j\mathbf k_n^T \mathbf r_m}$ is the candidate plane wave, propagating with wave number $\mathbf k_n \in \mathbb R^3$, to observation point $\mathbf r_m \in \mathbb R^3$, and $b_n(\omega) \in \mathbb C$ is the complex weight of the $n$th candidate plane wave.
The candidate plane waves can be obtained by sampling the wave number domain in a cubic grid.
Note that the eigenfunctions of the room used in Green's function can be expanded into a number of plane waves whose propagation directions in the wave number domain equals the characteristic frequency of the eigenfunction ($\|\mathbf k_n\|_2^2 = (\omega/c)^2$). This fact was used in \citep{Fernandez-grande2019} to regularize the sparse reconstruction problem as
\begin{equation}
  \underset{\mathbf b(\omega)}{\min} \quad \|\mathbf s(\omega) - \boldsymbol\Phi \mathbf b(\omega)\|_2 + \lambda\|\mathbf L(\omega)\mathbf b(\omega)\|_1
\end{equation}
where $\lambda \in \mathbb R^+$ and $\mathbf L(\omega) \in \mathbb R^{N\times N}$ is a diagonal matrix, where the diagonal elements express the distance between the characteristic frequency associated with the $n^\text{th}$ candidate plane wave and the angular excitation frequency $\omega$ as $| \|\mathbf k_N\|_2^2 - (\omega/c)^2 |$.

Note that the sparse reconstruction model is not directly comparable to the proposed sound field reconstruction. This is due to the sparse reconstruction relying on knowledge of the absolute locations of the microphone observations. The proposed algorithm, on the other hand, only requires the relative microphone locations on a unit-less observation grid.

\subsubsection{Experiments}
For the experiments, we use the simulated Listening Room from the previous section, with eight loudspeakers placed at the corners of the floor and halfway between the corners. We have two predefined zones in the middle of the room, which are bright and dark zone respectively. We now, sample random positions in the 32 by 32 x,y-grid 1~m above the floor and use those observations to estimate the RTFs within the zones. 

We compare the sparse reconstruction method to the deep learning-based model trained in the previous section.
Specifically, the room specialized models are used.

\begin{figure}
  \centering
    \fig{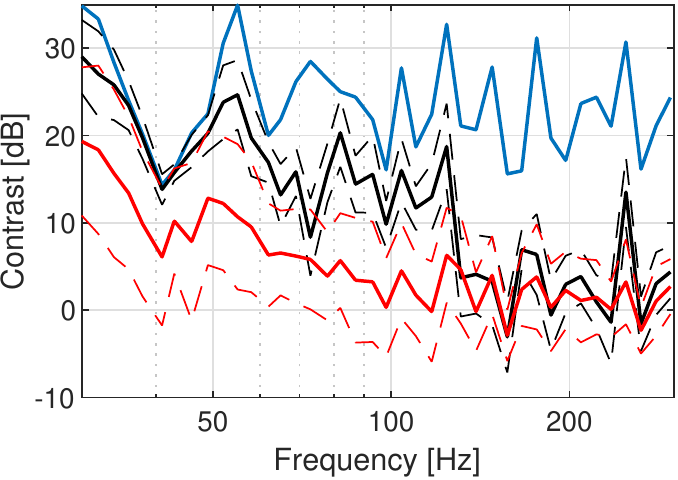}{7cm}{(a)}
    \fig{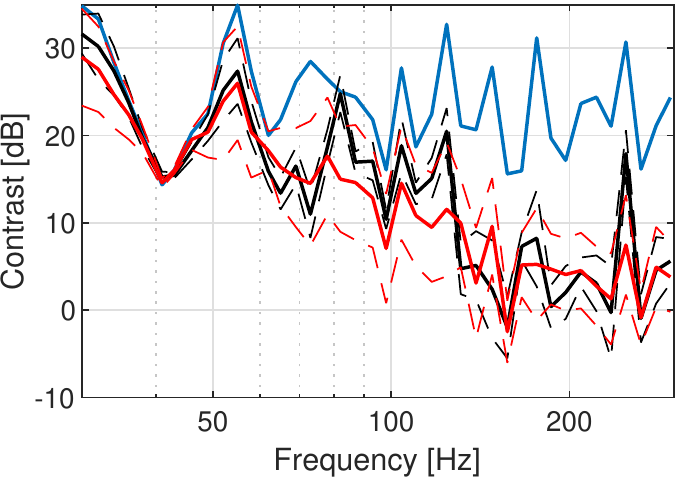}{7cm}{(b)}
  \caption{Contrast results for the dataset with no added uncertainty to the simulated Listening Room (50 different observation masks). (blue): Perfectly known TFs. (black): Deep learning model. (red): Sparse reconstruction. (dashed): $\pm 1$ standard deviation.}
  \label{fig:contrast_overfit}
\end{figure}

The resulting performance is evaluated in terms of the acoustic contrast over 50 random microphone samplings for each number of microphones. In Fig.~\ref{fig:contrast_overfit} the results are based on evaluations using the true RTFs when the loudspeaker weights are determined using either the true RTFs, estimated RTFs based on the model trained with simulated room with no added uncertainties, or estimates based on the sparse reconstruction. It is observed that the deep learning-based model performs better than the sparse reconstruction below 150~Hz for 5 and 15 microphones. Above 150~Hz, both models struggle to provide sufficiently accurate RTFs to create sound zones.

\begin{figure}
  \centering
    \fig{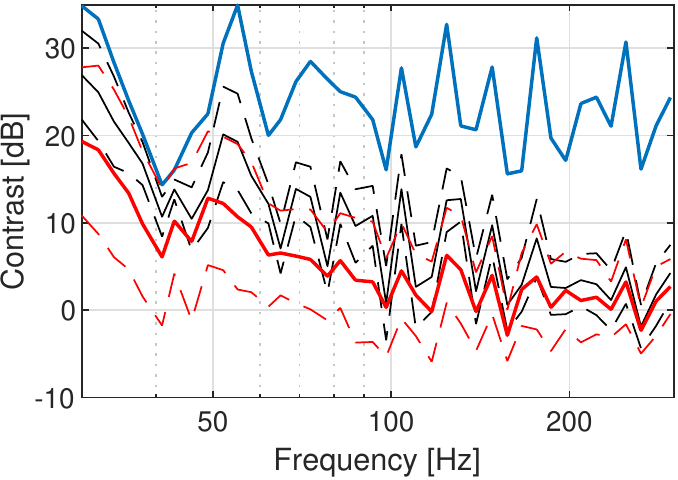}{7cm}{(a)}
    \fig{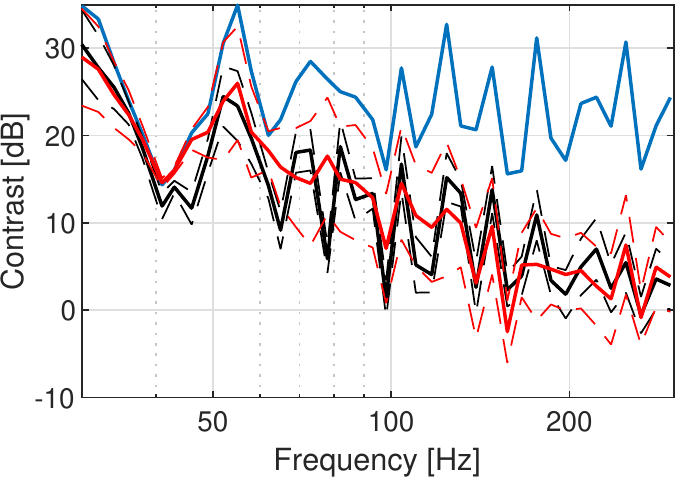}{7cm}{(b)}
  \caption{Contrast results for the simulated Listening Room with $l_x + \mathcal{U}(-1.0,1.0)$ m (50 different observation masks). (blue): Perfectly known TFs. (black): Deep learning model. (red): Sparse reconstruction. (dashed): $\pm 1$ standard deviation.}
  \label{fig:contrast_1m}
\end{figure}

In Fig.~\ref{fig:contrast_1m}, the model specialized for the Listening Room with $l_x + \mathcal{U}(-1.0,1.0)$ m, is compared to the sparse reconstruction. As expected, the resulting performance is reduced for the model. However, it is observed that there is still a benefit when using 5 microphones. At 15 microphones, on the other hand, the performance is comparable for both methods. 

These results indicate that sound zones could be created based on sound fields extrapolated from very few microphone positions. However, at this stage it requires models which are specialized to the particular room or a narrow range of rooms. Alternatively, it would be required to increase the number of microphones to improve the accuracy of the estimated RTFs.

\section{Conclusion}
In this paper, deep learning-based sound field reconstruction is evaluated using a new set of extensive measurements from real rooms, which are released alongside the paper.
The focus of the work is threefold: examine performance of simulation-based learning of magnitude reconstruction in real rooms, extend reconstruction to complex-valued sound fields, and show a sound zone application taking advantage of the reconstructed sound fields.
Experiments for each of the three directions indicate promising aspects of data-driven sound field reconstruction, even with a low number of arbitrarily placed microphones.

In the future, it would be of interest to investigate whether transfer learning can help bridge the discrepancies between simulated and real data.
With the addition of more rooms, some could be used in the training phase.
Furthermore, three-dimensional reconstruction can be achieved using available convolutional models designed specifically to solve three-dimensional problems.

\begin{acknowledgments}
    This work is part of the ISOBEL Grand Solutions project, and is supported in part by the Innovation Fund Denmark (IFD) under File No. 9069-00038A.
\end{acknowledgments}


\end{document}